\documentclass[twocolumn,showpacs,floatfix,pre,superscriptaddress]{revtex4-1}
\RequirePackage{lineno}

\usepackage{graphicx} 
\usepackage{dcolumn} 
\usepackage{epsf} 
\usepackage{amsmath}
\usepackage{bm} 
\usepackage{setspace}
\usepackage{color}
\usepackage[toc,page]{appendix}

\begin{document}


\title{Analysis of spatial correlations in a model 2D liquid through eigenvalues
and eigenvectors of atomic level stress matrices}

 \author{V.A.~Levashov}

 \affiliation{Technological Design Institute of Scientific Instrument Engineering,
 Novosibirsk, 630055, Russia}
 \affiliation{Department of Physics and Astronomy, University of 
Tennessee, Knoxville, TN 37996, USA}

 \author{M.G.~Stepanov}
 \affiliation{Department of Mathematics, University of Arizona, Tucson, 
AZ 85721, USA}


 \begin{abstract}
Considerations of local atomic level stresses associated 
with each atom represent a particular approach to address structures of 
disordered materials at the atomic level. We studied structural 
correlations in a two-dimensional model liquid using molecular dynamics 
simulations in the following way. We diagonalized the atomic level 
stress tensors of every atom and investigated correlations between the 
eigenvalues and orientations of the eigenvectors of different atoms as a 
function of distance between them. It is demonstrated that the suggested 
approach can be used to characterize structural correlations in 
disordered materials. In particular, we found that changes in the stress 
correlation functions on decrease of temperature are the most 
pronounced for the pairs of atoms with separation distance that 
corresponds to the first minimum in the pair density function. 
We also show that the angular dependencies of the stress correlation functions 
previously reported in [Phys. Rev.~E {\bf 91}, 032301 (2015)] related not to 
the alleged anisotropies of the Eshelby's stress fields, 
but to the rotational properties of the stress tensors.
 \end{abstract}

\pacs{64.70.Q--, 64.70.kj, 05.10.--a, 05.20.Jj}

\today

\maketitle


\section{Introduction}\label{s:intro}

It is relatively easy to describe structures of crystalline materials 
due to the presence of translational periodicity. This periodicity 
implies that atoms whose coordinates differ by a vector of translation 
have identical atomic environments. In glasses and liquids, in 
contrast, every atom, in principle, has a unique atomic environment 
\cite{EMa20111,ChenYQ2013}. Largely for this reason description of disordered 
materials continues to be a challenge. 
Many different approaches have been suggested to describe 
disordered structures. 
However, none of them allows establishing a 
clear link between the structural and dynamic properties of 
disordered matter \cite{EMa20111,ChenYQ2013}.

The concept of local atomic level stresses was introduced to describe 
model structures of metallic glasses and their liquids 
\cite{ChenYQ2013,Egami19801,Egami19821,Chen19881,Levashov2008B}. For a 
particle $i$ surrounded by 
particles $j$, with which it interacts through pair potential 
$U(r_{ij})$, the $\alpha \beta$ component of the atomic level stress tensor on 
atom $i$ is defined as 
\cite{Egami19801,Egami19821,Chen19881,Levashov2008B}:
\begin{eqnarray}
\sigma_i^{\alpha \beta}=\frac{1}{V_i}\sum_{j \ne i} 
\left[\frac{d U}{d r_{ij}}\right]\left(\frac{r_{ij}^{\alpha} r_{ij}^{\beta}}{r_{ij}}\right)\;\;\;.
\label{eq;stressdef1}
\end{eqnarray}
 The sum over $j$ in (\ref{eq;stressdef1}) is over all particles with 
which particle $i$ interacts. In (\ref{eq;stressdef1}) $V_i$ is the 
local atomic volume. By convention, the definition without $V_i$ 
corresponds to the local atomic level stress element 
\cite{Levashov2013,Levashov20111}. Note that $\alpha$-component of the force 
acting on particle $i$ from particle $j$ is 
$f^{\alpha}_{ij}=
\left[ d U(r_{ij})/d 
r_{ij}\right] \left(r_{ij}^{\alpha}/r_{ij}\right)$, 
where $\vec{r}_{ij} = \vec{r}_j - \vec{r}_i$ is the radius 
vector from $i$ to $j$. Also note that the atomic level stress tensor 
(\ref{eq;stressdef1}) is symmetric with respect to the indexes $\alpha$ and 
$\beta$. Thus in 3D it has 6 independent components \cite{Egami19821, 
Chen19881, Levashov2008B}, while in 2D it has 3 independent components.

There are several important results associated
with the concept of atomic level stresses. 
One result is the equipartition of the atomic level stress energies in 
liquids \cite{Egami19821,Chen19881,Levashov2008B}.
Thus the energies of the atomic level stress components were defined and it was demonstrated
for the studied model liquid systems in 3D that the energy of every stress component is equal to
$k_bT/4$. Thus the total stress energy, which is the sum of the energies of all six components,
is equal to $6 \cdot k_bT/4 = (3/2)k_bT$, i.e., the potential energy of a 
classical 3D harmonic oscillator.
An explanation for this result has been suggested \cite{Egami19821,Chen19881,Levashov2008B}. 
The equipartition breaks down in the glass state.
Then there was an attempt to describe glass transition and fragilities
of liquids on the basis of atomic level stresses \cite{Egami2007}.
Another result is related to the Green-Kubo expression for viscosity.
Thus the correlation function between the macroscopic stresses that
enter into the Green-Kubo expression for viscosity was decomposed into
the correlation functions between the atomic level stress elements. 
Considerations of the obtained atomic level correlation functions allowed 
demonstration of the relation between the propagation and dissipation 
of shear waves and viscosity. This result, after all, is not surprising
in view of the existing generalized 
hydrodynamics and mode-coupling theories \cite{HansenJP20061,Boon19911}.
However, in Ref.\cite{Levashov2013,Levashov20111,Levashov20141,Levashov2014B} the 
issue has been addressed from a new perspective and the relation
between viscosity and shear waves was demonstrated very explicitly.

Recently it has been claimed in Ref.\cite{Bin20151} that considerations of the 
correlations between the atomic level stresses allow observation of 
the angular dependent stress fields which are present in liquids in 
the absence of any external shear.
In many respects our attempt to understand the results presented in Ref.\cite{Bin20151} 
lead to the present publication. 

Here we demonstrate that the angular dependencies of the 
stress correlation functions presented in Ref.\cite{Bin20151} 
do not correspond to the angular dependent stress fields 
(which can exist in the system).
We show that the angular dependencies of the stress 
correlation functions observed in Ref.\cite{Bin20151}
originate from the rotational properties of the stress tensors.

However, the ideas presented here go beyond the scope of 
Ref.\cite{Bin20151}. Here we address the atomic level stresses and 
correlations between the atomic level stresses of atoms separated by some distance 
from a new and yet very natural perspective. 
It is surprising that this approach has not been investigated in detail 
before. Reasoning in a similar direction was presented in 
Ref.\cite{Kust2003a,Kust2003b}. However, considerations presented there 
do not address correlations between the atomic level stresses of 
different atoms.

This paper is organized as follows. In section \ref{sec:idea} 
the idea of the approach is presented. 
Section \ref{sec;tenstrans} is a reminder about transformational properties of the stress tensors. 
Atomic level stress correlations functions are discussed in the context of the present approach in section \ref{sec:Gcorrfunct}.
In section \ref{eq:eshelby1} the connection between the Eshelby's inclusion problem and the atomic level stress correlation
functions is analysed. 
In section \ref{sec:MDresults} the results of our MD simulations are described. 
We conclude in section \ref{sec:concl}.

\section{Stress tensor ellipses \label{sec:idea}}

\begin{figure}
\begin{center}
\includegraphics[angle=0,width=2.0in]{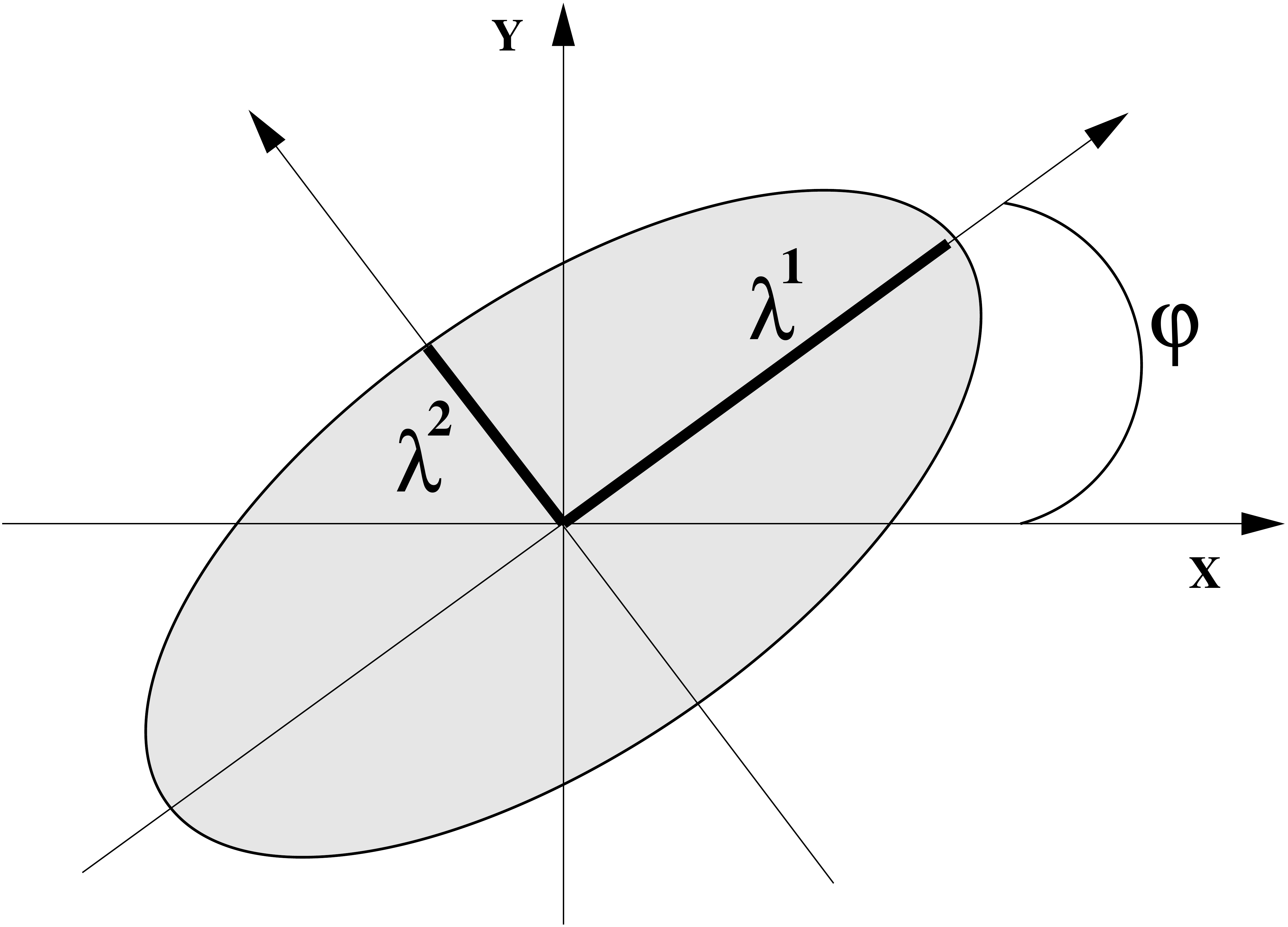}
\caption{Atomic level stress tensor of any atom can be diagonalized.
Obtained eigenvalues, $\lambda^1$ and $\lambda^2$, can be associated with the
lengths of the principal ellipse's axes. 
The orientation of the ellipse with respect to the reference coordinate system
is given by the angle, $\varphi$, between the longest ellipse's axis and the $x$-axis
of the reference frame. 
}\label{fig:1ellipes}
\end{center}
\end{figure}

The atomic level stress tensor $\sigma_i^{\alpha \beta}$ defined with equation 
(\ref{eq;stressdef1}) is real and symmetric.  Thus it can be 
diagonalised and, in 2D, two real eigenvalues ($\lambda^1_i$ and 
$\lambda^2_i$) and two real eigenvectors can be found.  
The tensor $\sigma_i^{\alpha \beta}$ in 2D has 3 independent components. 
These 3 parameters determine 2 eigenvalues and the 
rotation angle that describes orientation of the orthogonal eigenvectors 
with respect to the reference coordinate system.  Let us associate with 
each atom $i$ an ellipse with principal axes oriented along the 
eigenvectors of $\sigma_i^{\alpha \beta}$ and having lengths 
$\lambda^1_i$ and $\lambda^2_i$, as depicted in Fig.~\ref{fig:1ellipes}.

Previously atomic level stresses were discussed mostly in 3D.
In 3D symmetric atomic level stress tensors have 6 independent components. 
Thus, previously, in particular in discussions related to the atomic level stress energies,
it was assumed that the local atomic environment of an atom is described by 6 independent
stress components. However, in view of the present considerations, 
it is clear that if the atomic level stress tensor is diagonalized then its 
3 eigenvalues describe the geometry of local atomic
environment, while its 3 eigenvectors describe the orientation of the associated
ellipsoid with respect to the chosen coordinate system.

In model metallic glasses in 3D atoms often have 12 or 13 nearest 
neighbours \cite{EMa20111,ChenYQ2013,Levashov2008E}. Working with atomic 
level stresses effectively reduces the richness of all possible local 
atomic geometries to just $3$ numbers. Of course, that is more 
convenient than dealing with more numbers associated, for example, with 
the description based on Voronoi indexes \cite{EMa20111,ChenYQ2013}. 
However, it is unclear for which purposes it is enough to consider only 
3 numbers and for which purposes it may not be enough. In consideration 
of the stress correlations between two atoms in 3D there are 12 physically relevant 
parameters: 6 eigenvalues (3 on each atom) describe the geometries of 
the two ellipsoids and 6 parameters describe orientations of 
the ellipsoids with respect to the line from one ellipsoid to another.
A representation in a particular coordinate frame needs another 
3 parameters that describe the orientation of the line from one atom to another.

In Ref.\cite{Kust2003a,Kust2003b} correlations between the eigenvalues 
of the same atom has been considered for 2D and 3D Lennard-Jones 
liquids. There it was argued that there are correlations between the 
stress eigenvalues of the same atom.

\begin{figure}
\begin{center}
\includegraphics[angle=0,width=3.3in]{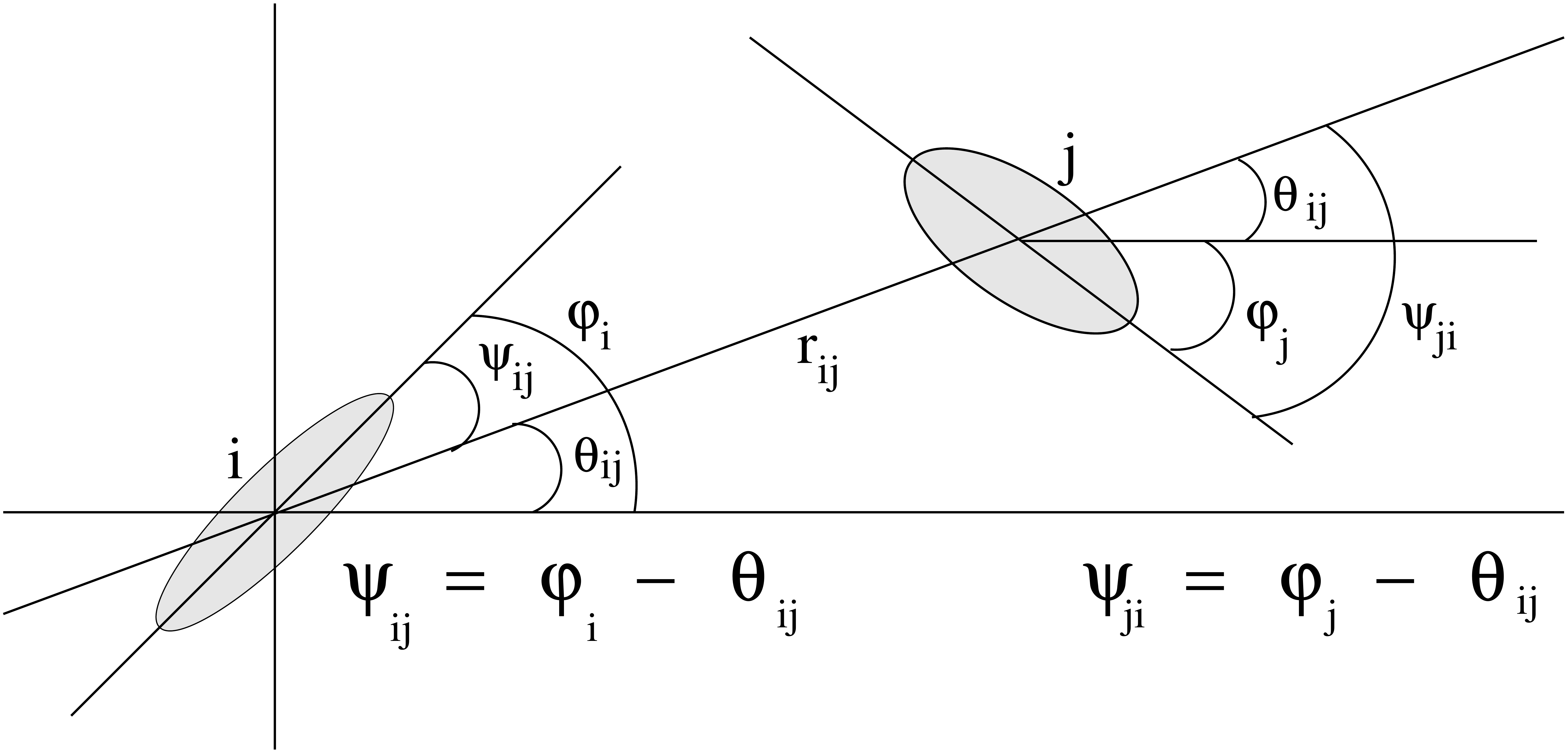}
\caption{The orientations of the ellipses with respect to the line connecting atoms
$i$ and $j$ are given by the angles $\psi_{ij}$ and $\psi_{ji}$. 
The orientation of the line connecting atoms $i$ and $j$ with respect to the 
$X$-axis of the reference frame is given by the angle $\theta_{ij}$.
}\label{fig:angles}
\end{center}
\end{figure}

Here we are interested in the correlations between the stress elements 
of different atoms. If we want to consider stress correlations between 
two different atoms in 2D then we associate ellipses with both atoms and 
consider the correlations between the eigenvalues and the orientations 
of the ellipses, see Fig.~\ref{fig:angles}. It is clear 
that in isotropic one component liquids all physically meaningful pair 
correlation functions should depend only on distance $r_{ij}$.

If the local atomic level stress tensor is known in the reference frame then its
eigenvalues and eigenvectors can be found. In 2D we have:
\begin{eqnarray}
&&\lambda^{1,2}_i \label{eq:lambda12}\\
&&= (1/2)\left[\left(\sigma_i^{xx}+\sigma_i^{yy}\right)
\pm \sqrt{\left(\sigma_i^{xx}-\sigma_i^{yy}\right)^2 + 4\left(\sigma_i^{xy}\right)^2}\right]\nonumber\\
&&\tan(\varphi_i^{1,2}) = (V^{1,2}_{i,y}/V^{1,2}_{i,x})=\sigma_i^{xy} / (\lambda^{1,2}_i - \sigma_i^{yy})\;\;. \label{eq;tanphi}
\end{eqnarray} 
Further we will assume that $\varphi_i \in (-\pi / 2, \pi / 2]$.
Physically angle $\varphi_i$ is defined up to an integer multiple 
of $\pi$ as rotation by angle $\pi$ does not change the ellipse.

In the orthogonal coordinate system based on the eigenvectors of a 
particular local atomic stress tensor this stress tensor is 
diagonal with eigenvalues $\lambda^1$ and $\lambda^2$ on the diagonal. 
In potentials with repulsive and attractive parts the values of some 
$\lambda$ can be negative. The negative value of $\lambda$ corresponds 
to the case when the atomic environment of an atom is dilated along the 
eigenvector associated with this $\lambda$. Here we assume that 
potentials that we consider are purely repulsive. Such systems held 
together at some density by periodic boundary conditions. In such cases, 
both $\lambda^1$ and $\lambda^2$ are positive and we order 
them to have $\lambda^1 \geq \lambda^2$.

\section{Transformations of stress tensors under rotations\label{sec;tenstrans}}

In this section we provide some well known facts about transformations 
of stress tensors under rotations \cite{Slaughter2002}. 
We will need these facts in our further considerations.

Let us suppose that there are $A$ and $B$ coordinate frames in 2D and that
frame $B$ is rotated with respect to frame $A$ on angle $\theta$
in the counterclockwise direction. The components of the stress tensor $S$
in frame $B$ can be expressed through the components of the stress 
tensor in frame $A$ using the rotation matrix $R(\theta)$:
\begin{eqnarray}
S_B = R(\theta) S_A R^T(\theta),\;\;\;\;
R(\theta)\equiv\begin{bmatrix} 
\cos(\theta) & \sin(\theta) \\ 
-\sin(\theta) & \cos(\theta) 
\end{bmatrix},\;\;\;
\label{SBfromSA}
\end{eqnarray}
where $R^T(\theta)$ is the transpose of $R(\theta)$.
In terms of components (\ref{SBfromSA}) leads to:
\begin{eqnarray}
&&\sigma_B^{xx}
= \sigma_A^{xx}\left[\cos(\theta)\right]^2 
+ \sigma_A^{yy}\left[\sin(\theta)\right]^2
+\sigma_A^{xy}\sin(2\theta)\;,\label{eq:rotABxx}\\
&&\sigma_B^{yy}
= \sigma_A^{xx}\left[\sin(\theta)\right]^2 
+ \sigma_A^{yy}\left[\cos(\theta)\right]^2
-\sigma_A^{xy}\sin(2\theta)\;,\label{eq:rotAByy}\\
&&\sigma_B^{xy}
= -(1/2)\left[\sigma_A^{xx}-\sigma_A^{yy}\right]\sin(2\theta) 
+ \sigma_A^{xy}\cos(2\theta)\;.\label{eq:rotABxy}
\end{eqnarray}

Let us now suppose that the angle between the first eigenvector of atom $i$ 
and the $\hat{x}$-axis of our reference frame is $\varphi_i$.
In the frame of its eigenvectors the components of the stress tensor of atom $i$
are $\sigma_i^{xx}=\lambda_i^1$, $\sigma_i^{yy}=\lambda_i^2$, 
$\sigma_i^{xy}=0$, $\sigma_i^{yx}=0$. In order to find the components of the
stress tensor of atom $i$ in our reference frame we should ``rotate" the components
of the stress tensor in the frame of its eigenvectors on angle $-\varphi_i$ using 
(\ref{eq:rotABxx},\ref{eq:rotAByy},\ref{eq:rotABxy}). Thus we get:
\begin{eqnarray}
&&\sigma^{xx}_i= \lambda_i^1\cos^2(\varphi_i) 
+ \lambda_i^2\sin^2(\varphi_i)\;,\label{eq:rotixx}\\
&&\sigma^{yy}_i= \lambda_i^1\sin^2(\varphi_i) 
+ \lambda_i^2\cos^2(\varphi_i)\;,\label{eq:rotiyy}\\
&&\sigma^{xy}_i= (1/2)\left(\lambda_i^1-\lambda_i^2\right)\sin(2\varphi_i)\;.\label{eq:rotixy}
\end{eqnarray}
Note also that: $\sigma_i^{xx}-\sigma_i^{yy} = 
\left(\lambda_i^1-\lambda_i^2\right)\cos(2\psi_{ij})$.

\section{Correlation functions between the elements of atomic level stress tensors of
different atoms\label{sec:Gcorrfunct}}

In this section we derive the expressions for selected correlation functions between the atomic 
level stress elements in terms of eigenvalues and eigenvectors of atomic level stress matrices.

\subsection{Correlation functions in the directional frame \label{sec:dirframe}} 

It is useful to start this section from an argument which plays a very important
role in this paper.

Let us consider a pair of atoms $i$ and $j$ separated by radius 
vector $\bm{r}_{ij}=\bm{r}_j-\bm{r}_i$.
We associate with the direction of $\bm{r}_{ij}$ a directional 
coordinate ``$\bm{r}_{ij}$-frame" whose $\bm{\hat{x}}$-axis is along $\bm{r}_{ij}$. 
The notations $\sigma_{ij}^{\alpha\beta}(i)$ and $\sigma_{ij}^{\delta\gamma}(j)$ will 
be used for the $\alpha\beta$ and $\gamma\delta$ components of the stress tensors of atoms $i$ 
and $j$ in the $\bm{r}_{ij}$-frame.
Further we consider the products $\sigma_{ij}^{\alpha\beta}(i)\sigma_{ij}^{\delta\gamma}(j)$  
in the $\bm{r}_{ij}$-directional frame and
average such products over the pairs of atom separated by radius vector $\bm{r}_{ij}=\bm{r}$.
\emph{It is important that this averaging is performed over the values of the stress tensor components in the
representation associated with the $\bm{r}_{ij}$-frame.}

For the following it is necessary to realize that for isotropic systems of particles the averaging,
\begin{eqnarray}
\langle \sigma_{ij}^{\alpha\beta}(i)\sigma_{ij}^{\delta\gamma}(j)\rangle _{\bm{r}_{ij}=\bm{r}}\;,\;\;
\label{eq:isotropisity1} 
\end{eqnarray}
should not depend on the direction of $\bm{r}$, while it can depend on $r=|\bm{r}|$.
This is essentially what isotropicity means. 

\subsection{Transformation of correlation functions under rotations\label{sec:corrrotate}}

Our goal in this section is to express the correlation functions between the stress tensor components
in an arbitrary frame in terms of the correlation functions in the $\bm{r}_{ij}$-frame introduced in the
previous subsection.

Thus, let us express the product $\sigma^{xy}(i)\sigma^{xy}(j)$ 
in the coordinate frame which is rotated on the angle $-\theta_{ij}$ 
with respect to $\bm{r}_{ij}$-frame
in terms of stress tensor components in the $\bm{r}_{ij}$-frame.

For this we should rotate, according to (\ref{eq:rotABxx},\ref{eq:rotAByy},\ref{eq:rotABxy}),
the stress tensor components of atoms $i$ and $j$ in the $\bm{r}_{ij}$-frame on the angle $-\theta_{ij}$ 
and then form the products of the stress tensor components in the rotated frame. 
From (\ref{eq:rotABxy}) we get:
\begin{eqnarray}
&&\sigma_{\theta_{ij}}^{xy}(i)\sigma_{\theta_{ij}}^{xy}(j)\label{eq:try-sxy-sxy-1}\\
&&=(1/4)\left[\sigma_{ij}^{xx}(i)-\sigma_{ij}^{xx}(i)\right]
\left[\sigma_{ij}^{xx}(j)-\sigma_{ij}^{xx}(j)\right]\left[\sin(2\theta_{ij})\right]^2\nonumber\\
&&+(1/4)\left[\sigma_{ij}^{xx}(i)-\sigma_{ij}^{xx}(i)\right]\sigma_{ij}^{xy}(j)
\left[\sin(4\theta_{ij})\right]\nonumber\\
&&+(1/4)\sigma_{ij}^{xy}(i)\left[\sigma_{ij}^{xx}(j)-\sigma_{ij}^{xx}(j)\right]
\left[\sin(4\theta_{ij}))\right]\nonumber\\
&&+\sigma_{ij}^{xy}(i)\sigma_{ij}^{xy}(j)
\left[\cos(2\theta_{ij})\right]^2\;.\nonumber
\end{eqnarray}
The right hand side of (\ref{eq:try-sxy-sxy-1}) can be expanded in terms containing the products
of the stress tensor components in the $\bm{r}_{ij}$-frame.

Let us now average (\ref{eq:try-sxy-sxy-1}) over the pairs of atoms $i$ and $j$ 
separated by $\bm{r}_{ij}=\bm{r}$ (this fixes the value of $\theta_{ij}=\theta$).
For briefness and as an example let us consider a particular term, 
$\sigma_{ij}^{xy}(i)\sigma_{ij}^{xy}(j)\left[\cos(2\theta_{ij})\right]^2$,
that appears on the right hand side of (\ref{eq:try-sxy-sxy-1}). 
In performing the averaging we get:
\begin{eqnarray}
&&\langle \sigma_{ij}^{xy}(i)\sigma_{ij}^{xy}(j)\left[\cos(2\theta_{ij})\right]^2\rangle _{\bm{r}_{ij}=\bm{r}}
\label{eq:termexample1}\\
&&=\langle \sigma_{ij}^{xy}(i)\sigma_{ij}^{xy}(j)\rangle _{\bm{r}_{ij}=\bm{r}}\left[\cos(2\theta)\right]^2\;.\;\;\;\label{eq:termexample2}
\end{eqnarray}
In the transition from (\ref{eq:termexample1}) to (\ref{eq:termexample2}) 
the $\left[\cos(2\theta_{ij})\right]^2$ was taken out of the averaging since
the averaging is performed for a fixed value of $\bm{r}_{ij}=\bm{r}$ and it also means that the averaging is performed for
a fixed value $\theta_{ij}=\theta$. It follows from the previous subsection (\ref{sec:dirframe}) that 
in isotropic medium $\langle \sigma_{ij}^{xy}(i)\sigma_{ij}^{xy}(j)\rangle _{\bm{r}_{ij}=\bm{r}}$ should not depend 
on the direction of $\bm{r}$, but can depend on $r=|\bm{r}|$.

Thus, in performing the averaging of the products of the stress tensor components 
in (\ref{eq:try-sxy-sxy-1}), as it was done in (\ref{eq:termexample1},\ref{eq:termexample2}),
it is possible to average over all pairs of atoms separated by $r_{ij}=r$ irrespectively of the direction
of $\bm{r}$. \emph{It is only necessary to ensure that the values of the stress tensor components
on the right hand side of (\ref{eq:try-sxy-sxy-1}) are always calculated in 
the directional $\bm{r}_{ij}$-frame corresponding to each pair of atoms $i$ and $j$.}

It follows from the above considerations that the value of the correlation function 
$\langle  \sigma^{xy}(i)\sigma^{xy}(j)\rangle _{\bm{r}_{ij}=\bm{r}}$ at some $r$ and $\theta$
can be expressed as a linear combination of the correlation functions between the atomic level stress elements
in the $\bm{r}_{ij}$-frame multiplied on some functions of $\theta$.
Note that the dependence on $\theta$ in (\ref{eq:try-sxy-sxy-1}) appears in the result
of rotation from the $\bm{r}_{ij}$-frame into the frame in which $\bm{r}_{ij}$ forms angle
$\theta$ with the $\hat{x}$-axis.
Thus in an isotropic medium the physical essence of 
the atomic level stress correlations is contained in the correlation functions 
associated with the $\bm{r}_{ij}$-frame. 
In an isotropic medium these correlations function should depend only on distance.

\subsection{Expressions for the selected stress correlation functions in terms of eigenvalues and eigenvectors in the
$\bm{r}_{ij}$-directional frame\label{sec:stresscorr}}

It follows from the two previous subsections (\ref{sec:dirframe}, \ref{sec:corrrotate}) that in order
to find correlation functions of the atomic level stress components in any coordinate frame it is sufficient
to know correlation functions in the directional $\bm{r}_{ij}$-frame. 

It is easy to express the correlation
functions in the $\bm{r}_{ij}$-frame in terms of eigenvalues and eigenvectors of the atomic level stress matrices.
Let us suppose that the first eigenvectors of the stress matrices of atoms $i$ and $j$ form angles
$\psi_{ij}$ and $\psi_{ji}$ with the direction $\bm{r}_{ij}$, as shown in Fig.\ref{fig:angles}.

From (\ref{eq:rotixx},\ref{eq:rotiyy}) it follows that
the rotation invariant atomic level pressure on atom $i$ is:
\begin{eqnarray}
p_i \equiv (1/2)\left[\sigma_{ij}^{xx}(i)+\sigma_{ij}^{yy}(i)\right]=(1/2)\left(\lambda_i^1 + \lambda_i^2\right)\;.\;\;\label{eq:pidefine}
\end{eqnarray}
Correspondingly
\begin{eqnarray}
\langle p_ip_j\rangle _{\bm{r}_{ij}=\bm{r}}=
(1/4)\langle \left(\lambda_i^1 + \lambda_i^2\right)\left(\lambda_j^1+\lambda_j^2\right)\rangle _{\bm{r}_{ij}=\bm{r}}\;.\label{eq:pipj01}
\end{eqnarray}
It also follows from (\ref{eq:rotixx},\ref{eq:rotiyy},\ref{eq:rotixy}) that:
\begin{eqnarray}
&&\langle p_i\sigma_{ij}^{xy}(j)\rangle _{\bm{r}_{ij}=\bm{r}}\label{eq:pisxyj01}\\
&&=(1/4)\langle \left(\lambda_i^1 + \lambda_i^2\right)
\left(\lambda_j^1-\lambda_j^2\right)\sin(2\psi_{ji})\rangle _{\bm{r}_{ij}=\bm{r}}\;,\nonumber\\
&&\langle p_i\left[\sigma_{ij}^{xx}(j)-\sigma_{ij}^{yy}(j)\right]\rangle _{\bm{r}_{ij}=\bm{r}}\label{eq:pisdiffj01}\\
&&=(1/2)\langle \left(\lambda_i^1 + \lambda_i^2\right)
\left(\lambda_j^1-\lambda_j^2\right)\cos(2\psi_{ji})\rangle _{\bm{r}_{ij}=\bm{r}}\;,\nonumber\\
&&\langle \sigma_{ij}^{xy}(i)\sigma_{ij}^{xy}(j)\rangle _{\bm{r}_{ij}=\bm{r}}\label{sxyisxyj01}\\
&&= (1/4)\langle \left(\lambda_i^1-\lambda_i^2\right)\left(\lambda_j^1-\lambda_j^2\right)
\left[\sin(2\psi_{ij})\sin(2\psi_{ji})\right]\rangle _{\bm{r}_{ij}=\bm{r}}\;,\;\;\nonumber\\
&&\langle \sigma_{ij}^{xy}(i)\left[\sigma_{ij}^{xx}(j)-\sigma_{ij}^{yy}(j)\right]\rangle _{\bm{r}_{ij}=\bm{r}}\label{eq:sxyisdiffj01}\\
&&= (1/2)\langle \left(\lambda_i^1-\lambda_i^2\right)\left(\lambda_j^1-\lambda_j^2\right)
\left[\sin(2\psi_{ij})\cos(2\psi_{ji})\right]\rangle _{\bm{r}_{ij}=\bm{r}}\;,\;\;\nonumber\\
&&\langle \left[\sigma_{ij}^{xx}(i)-\sigma_{ij}^{yy}(i)\right]
\left[\sigma_{ij}^{xx}(j)-\sigma_{ij}^{yy}(j)\right]\rangle _{\bm{r}_{ij}=\bm{r}}\label{eq:sdiffisdiffj01}\\
&&= \langle \left(\lambda_i^1-\lambda_i^2\right)\left(\lambda_j^1-\lambda_j^2\right)
\left[\cos(2\psi_{ij})\cos(2\psi_{ji})\right]\rangle _{\bm{r}_{ij}=\bm{r}}\;,\;\;\nonumber
\end{eqnarray}
Note that the right hand sides of 
(\ref{eq:pidefine},\ref{eq:pipj01},\ref{eq:pisxyj01},\ref{eq:pisdiffj01},\ref{sxyisxyj01},\ref{eq:sxyisdiffj01},\ref{eq:sdiffisdiffj01})
depend on the invariant parameters of the atomic level stress ellipses and on their
rotation invariant orientations with respect to the direction of $\bm{r}_{ij}$.
Thus, in finding how
(\ref{eq:pipj01},\ref{eq:pisxyj01},\ref{eq:pisdiffj01},\ref{sxyisxyj01},\ref{eq:sxyisdiffj01},\ref{eq:sdiffisdiffj01})
depend on $r$ in isotropic systems it is possible to average over all pairs
separated by $r$ irrespective of the orientation of $\bm{r}$.
This is in agreement with the argument from subsection (\ref{sec:dirframe}) that states that correlation functions
between the components of atomic level stresses in the directional $\bm{r}_{ij}=\bm{r}$ frame should not depend on the direction
of $\bm{r}$.

\subsection{The stress correlation function $\langle \sigma^{xy}(i)\sigma^{xy}(j)\rangle _{\bm{r}_{ij}=\bm{r}}$ 
in the arbitrary reference frame
expressed in terms of eigenvalues and eigenvectors.
\label{ssec:corrsxyisxyj}}

The correlation function $\langle \sigma^{xy}(i)\sigma^{xy}(j)\rangle _{\bm{r}_{ij}=\bm{r}}$ 
in any fixed reference frame depends on
$\bm{r}$, i.e., on $r$ and $\theta$.
Using expressions (\ref{eq:try-sxy-sxy-1},\ref{eq:termexample1},\ref{eq:termexample2},\ref{sxyisxyj01},\ref{eq:sxyisdiffj01},\ref{eq:sdiffisdiffj01})
it is straightforward (although a bit tedious) to obtain the following expression:
\begin{eqnarray}
&&\langle \sigma^{xy}(i)\sigma^{xy}(j)\rangle_{\bm{r}_{ij}=\bm{r}}\label{eq:sixysjxyref02}\\
&& = (1/8) \left[F_1(r) 
- F_2(r)\cos(4\theta) + F_3(r)\sin(4\theta) \right],\nonumber
\end{eqnarray} 
where
\begin{eqnarray}
&&F_1(r) \equiv \langle\mathcal{F}_1\rangle_{\bm{r}_{ij}=r}\label{eq;F1}\\
&&=\langle (\lambda^1_i - \lambda^2_i)(\lambda^1_j - 
\lambda^2_j) \cos(2\psi_{ij}-2\psi_{ji})\rangle _{\bm{r}_{ij}=\bm{r}}\;,\;\;\;\nonumber\\
&&F_2(r) \equiv \langle\mathcal{F}_2\rangle_{r_{ij}=r}\label{eq;F2}\\
&&= \langle (\lambda^1_i - \lambda^2_i)(\lambda^1_j - 
\lambda^2_j) \cos(2\psi_{ij}+2\psi_{ji})\rangle _{\bm{r}_{ij}=\bm{r}}\;,\;\;\;\nonumber\\
&&F_3(r)\equiv\langle \mathcal{F}_3\rangle_{\bm{r}_{ij}=\bm{r}} \label{eq;F3}\\
&&= \langle (\lambda^1_i - \lambda^2_i)(\lambda^1_j - 
\lambda^2_j) \sin(2\psi_{ij}+2\psi_{ji})\rangle _{\bm{r}_{ij}=\bm{r}}\;.\;\;\;\nonumber
\end{eqnarray}
Note that dependence of (\ref{eq:sixysjxyref02}) on $\theta$ originates from (\ref{eq:try-sxy-sxy-1}),
i.e., from the rotation from the directional $\bm{r}_{ij}$-frame into the coordinate frame that forms
angle $\theta$ with the direction of $\bm{r}_{ij}=\bm{r}=[r\cos(\theta),r\sin(\theta]$. Thus the dependence of (\ref{eq:sixysjxyref02}) on
$\theta$ merely reflects the rotational properties of the stress tensors. 
Also note that all physically meaningful information about correlations between the parameters
of atomic level stresses is contained in functions $F_1(r)$, $F_2(r)$, and $F_3(r)$.

In finding $F_1(r)$, $F_2(r)$, and $F_3(r)$ in isotropic medium 
the averaging can be performed over all pairs
of atoms $i$ and $j$ separated by distance $r$ irrespectively of the direction 
of $\bm{r}$.

Two alternative derivations of the formulas (\ref{eq:sixysjxyref02},\ref{eq;F1},\ref{eq;F2},\ref{eq;F3})
are presented in Appendices (\ref{sec:mishapdx}, \ref{sec:derivation3})

{\em In order to understand the meaning of correlation function $F_1$ let us consider
the contribution from some atoms $i$ and $j$ to this function. It follows from
(\ref{eq;F1}) that:}\\
1) If one of the ellipses is a circle, for example $\lambda_i^1 = \lambda_i^2$, then the
contribution from this pair of atoms is zero. Thus correlation function $F_1(r)$ contains contributions 
only from those pairs of atoms in which there are finite shear 
deformations of the environments of both atoms.\\ 
2) If ellipses of atoms $i$ and $j$ have the same orientation 
with respect to the line connecting them then
$\cos(2\psi_{ij}-2\psi_{ji})=1$ and the contribution 
from this pair of ellipses is the maximum possible contribution
from the pairs of ellipses with the same distortions. \\
3) If ellipses of atoms $i$ and $j$ are orthogonal to each 
other, i.e., $\psi_{ij} = \psi_{ji} \pm \pi/2$ then $\cos(2\psi_{ij}-2\psi_{ji})=-1$ and
the contribution from this pair is the minimum possible contribution.\\
4) If $\psi_{ij} = \psi_{ji} \pm \pi/4$ then the contribution is zero.\\

Note also the following.
If large axes of the ellipses of atoms $i$ and $j$ are aligned then 
these ellipses have the same orientation with respect to any line, 
not only the line that connects them. Thus it is likely that rather simple
organization of ellipses provides a maximum to the function $F_1$. 
It is the organization when all ellipses have the same shear distortions
and the same orientations. This observation might be of interest for understanding
the nature of viscosity. It follows from the Green-Kubo expression that viscosity is determined
by decay in time of the function $F_1(r)$, i.e., for calculations of viscosity
it is necessary to consider stress of atom $i$ at time zero and stress of 
atom $j$ at time $t$ ($F_2(r)$ does not contribute since 
integration over $\theta$ in (\ref{eq:sixysjxyref02}) leads to zero). \\ 

{\em In order to understand the meaning of correlation function $F_2(r)$ 
from (\ref{eq;F2}) note the following:}\\
1) As in the case with $F_1(r)$, only pairs of atoms in which both atoms have shear distortions contribute.\\
2) The maximum contribution, for the given distortions, comes from the ellipses for which
$\psi_{ij} =- \psi_{ji}$, i.e., from those ellipses 
whose orientations are mirror-symmetric with respect to the line connecting them.\\
3) If the deviation from the mirror symmetry is $\pi/2$, 
i.e., $\psi_{ij} =- \psi_{ji} \pm \pi/2$ then the contribution is the minimum possible contribution.\\
4) If $\psi_{ij} = -\psi_{ji} \pm \pi/4$ then the contribution is zero.\\

{\em Due to a mirror symmetry we must have $F_3(r) = 0$}. 
This is because 
reflection with respect to the direction from $i$ to $j$ changes the 
signs of angles $\psi_{ij}$ and $\psi_{ji}$, but does not change the 
eigenvalues. In our simulations $F_3(r)$ averages to zero up to the noise 
level.\\

\subsection{Stress correlation function $\langle p_i\sigma^{xy}(j)\rangle _{\bm{r}_{ij}=\bm{r}}$}

From (\ref{eq:rotABxy},\ref{eq:pidefine},\ref{eq:pisxyj01},\ref{eq:pisdiffj01}),
similarly to how it was done for $\langle \sigma^{xy}(i)\sigma^{xy}(j)\rangle _{\bm{r}_{ij}=\bm{r}}$, 
we get:
\begin{eqnarray}
&&\langle p_i \sigma_j^{xy}\rangle _{\bm{r}_{ij}=\bm{r}}\label{eq;pxy1}\\
&&=(1/4)\left[F_4(r)\cos(2\theta)+F_5(r)\sin(2\theta)\right]\;\;,\;\;\nonumber
\end{eqnarray}
where
 \begin{eqnarray}
   F_4(r) &=& \langle  (\lambda^1_i+\lambda^2_i) (\lambda^1_j-\lambda^2_j) 
\sin(2\psi_{ji}) \rangle _{\bm{r}_{ij}=\bm{r}}\;\;,\;\;\\
   F_5(r) &=& \langle  (\lambda^1_i+\lambda^2_i) (\lambda^1_j-\lambda^2_j) 
\cos(2\psi_{ji})\rangle _{\bm{r}_{ij}=\bm{r}}\;\;.\;\; \label{eq;F5}
 \end{eqnarray}
In finding $F_4(r)$ and $F_5(r)$ in isotropic medium 
the averaging can be performed over all pairs
of atoms $i$ and $j$ separated by distance $r$ irrespectively of the direction 
of $\bm{r}$.
 
Due to mirror symmetry, the function $F_4$ should average to zero (it does in simulations).\\

{\em In order to understand the meaning of $F_5$ from (\ref{eq;F5}) note the following:}\\
1) The larger is the pressure on atom $i$ and the shear distortion of atom $j$,
the larger is the contribution from this pair to $F_5$.\\ 
2) If the ellipse of atom $j$ is aligned with the direction from $i$ to $j$ 
then $\cos(2\psi_{ji})=1$ and there is the maximum possible contribution for the given ellipses' shapes.\\ 
3) If the ellipse of atom $j$ is orthogonal to the direction from $i$ to $j$ 
then $\cos(2\psi_{ji})=-1$ and there is the minimum possible contribution for the given ellipses' shapes.\\
4) If $\psi_{ji}=\pi/4$ then the contribution is zero.

\subsection{Stress correlation function $\langle (\sigma^{xx}_i-\sigma^{yy}_i)\sigma_j^{xy}\rangle _{\bm{r}_{ij}=\bm{r}}$}

From (\ref{eq:rotABxx},\ref{eq:rotAByy},\ref{eq:rotABxy}) and 
(\ref{sxyisxyj01},\ref{eq:sxyisdiffj01},\ref{eq:sdiffisdiffj01}),
similarly to how it was done for $\langle \sigma_i^{xy}\sigma_j^{xy}\rangle _{\bm{r}_{ij}=\bm{r}}$, we get:
\begin{eqnarray}
&&\langle (\sigma^{xx}_i-\sigma^{yy}_i)\sigma_j^{xy}\rangle _{\bm{r}_{ij}=\bm{r}} \label{eq:sxysdiffcorrf01}\\
&&=(1/2)F_1(r)\sin(4\theta) + F_6(r) \cos(4\theta)\;,\nonumber 
\end{eqnarray}
where  $F_1(r)$ is given by expression (\ref{eq;F1}) and: 
\begin{eqnarray}
F_6(r) = \langle (\lambda^1_i-\lambda^2_i)(\lambda^1_j-\lambda^2_j)
\cos(2\psi_{ij})\sin(2\psi_{ji})\rangle _{\bm{r}_{ij}=\bm{r}}.\;\;\;\;\;\;\label{eq:F6lambda}
\end{eqnarray}
In finding $F_6(r)$ in an isotropic medium the averaging can be performed over all pairs
of atoms $i$ and $j$ separated by distance $r$ irrespective of the direction 
of $\bm{r}$.

The function $F_6$ should average to zero due to mirror symmetry with respect
to the direction from $i$ to $j$ since under reflection
$\cos(\psi_{ij})$ does not change sign, while $\sin(\psi_{ji})$ does.
We verified this in our simulations.

\subsection{Simpler correlation functions and normalization of the correlation functions}

Correlation functions $F_{1,2,3,4,5,6}$ are 
somewhat complicated as they represent averages over three or four parameters. 
Before considering them it makes sense to consider 
simpler correlation functions which represent averaged products on two parameters only. 
It is expectable that stresses of particles which are far away from each other are not correlated. 
This makes it reasonable to consider the following correlation functions:
 \begin{eqnarray}
   && G_{pp}(r) = 
(1/Z_{+}^2)\langle (\lambda^1_i+\lambda^2_i)(\lambda^1_j+\lambda^2_j)\rangle _{r_{ij}=r} 
- 1\;\;,\;\;\label{eq;Gpp}\\
  && G_{mm}(r) = 
(1/Z_{-}^2)\langle (\lambda^1_i-\lambda^2_i)(\lambda^1_j-\lambda^2_j)\rangle _{r_{ij}=r} 
- 1\;\;,\;\;\label{eq;Gmm}\\
   && G_{mp}(r) = (1/Z_{+} 
Z_{-})\langle (\lambda^1_i-\lambda^2_i)(\lambda^1_j+\lambda^2_j)\rangle _{r_{ij}=r} - 
1\;\;,\;\;\label{eq;Gmp}\\
&&C_{2\pm}(r) = \langle  \cos(2\psi_{ij} \pm 2\psi_{ji}) \rangle _{r_{ij}=r}\;\;,\;\;
\label{eq;Cpm}
\end{eqnarray}
where $Z_\pm = \langle  \lambda^1_i \pm \lambda^2_i \rangle$. 

Functions $G_{pp}(r)$, $G_{mm}(r)$, and $G_{mp}(r)$ describe correlations
between the eigenvalues (or eigenstresses) of the stress matrices of atoms
$i$ and $j$ without taking into account the orientations of the eigenvectors. 
Note that since $p_i = (1/2)(\lambda_i^1+\lambda_i^2)$ the function $G_{pp}(r)$ 
from (\ref{eq;Gpp}) is directly related to the pressure-pressure correlation 
function between atoms $i$ and $j$. 
It follows from Appendix \ref{sec:mishapdx} and formula 
(\ref{eq:lambda12}) that the function $G_{mm}$ represents
correlations between the total amounts of shear on atoms $i$ and $j$.
Finally, $G_{mp}(r)$ describes correlations between the total shear on atom $i$ and the
total pressure on atom $j$.
Functions $C_{2\pm}(r)$ from (\ref{eq;Cpm}) describe correlations in the orientations of the eigenvectors
of the stress matrices of atoms $i$ and $j$ without taking into account the magnitudes of the
eigenvalues.

It is also reasonable to introduce normalized versions of the correlation functions 
$F_{1,2,3,4,5,6}$:
 \begin{eqnarray}
   && \tilde{F}_{1,2,3,6}(r_{ij}) \equiv 
F_{1,2,3,6}/Z_{-}^2\;\;\;,\;\;\label{eq;tF1}\\
   && \tilde{F}_{4,5} \equiv 
F_{4,5}/(Z_{+}Z_{-})\;\;\;.\;\;\label{eq;tF2}
 \end{eqnarray}

\section{Analogy with the Eshelby's inclusion problem \label{eq:eshelby1}}

In this section we discuss from the perspective of the Eshelby's 
inclusion problem \cite{Eshelby1957,Eshelby1959,Slaughter2002,Weinberger2005,Dasgupta2013}
the stress correlation function which is analogous to the
atomic level stress correlation function $\langle \sigma_i^{xy}\sigma_j^{xy}\rangle _{\bm{r}_{ij}=\bm{r}}$
discussed in the previous section.
In drawing this analogy it is assumed that the central atom $i$ is analogous to the Eshelby's inclusion ($I$)
that generates a stress field in the matrix at point $J$ (on atom $j$).
In particular, we argue that the angular dependence of 
the $\langle \sigma_i^{xy}\sigma_j^{xy}\rangle _{\bm{r}=\bm{r}}$ stress correlation function 
obtained in Ref.\cite{Bin20151} is related to the
rotational properties of the stress tensors and not to the
anisotropy of the stress field associated with the Eshelby's solution.

There are two points which we need from the the Eshelby's solution. 
1) The final strain and stress fields in the inclusion after the deformation, 
placing the inclusion back into the matrix, and joining are constant. 
The final strain and stress fields in the inclusion, of course, 
depend on the unconstrained strain initially applied to the inclusion. 
2) If we know the unconstrained strain applied to the inclusion then the final
strain and stress fields in the inclusion and in the matrix can be found. 
Further we assume that there is a one-to-one correspondence 
between the stress fields in the inclusion and in the matrix.
See also Appendix \ref{app:inclusion}.

We are interested in the correlation functions between the inclusion ($I$) 
and some point ($J$) in the matrix. 
Similarly to how it was done for the atomic level stresses, 
we can associate with $I$ the stress ellipse whose 
parameters, ($\lambda^1_I$, $\lambda^2_I$), and whose orientation, 
$\psi_{IJ}$, with respect to $\bm{r}_{IJ}$ are known. 
The fact that the stress field is the same everywhere in the inclusion serves well
for this purpose. 

Since the inclusion's stress ellipse is known,  
the stress field at any point $J$ in the matrix can be found.  
Since the stress tensor at point $J$ is known it can be diagonalized 
and thus it is possible to associate with point $J$ its own stress ellipse with 
parameters ($\lambda^1_J$, $\lambda^2_J$) and the orientation
$\psi_{JI}$ with respect to $\bm{r}_{IJ}$. 

At this point it becomes apparent that considerations of correlations for the 
Eshelby's inclusion problem are quite similar to the considerations that were already done 
for the atomic level stresses. There is, however, an important difference.
Thus, in the case of atomic level stresses correlations between the parameters of atomic level
stress ellipses have a \emph{probabilistic} character.
In contrast, in the case of the Eshelby's inclusion problem the stress field in the inclusion
\emph{deterministically defines} the stress field in the matrix.
Thus, $\lambda_J^{1}$, $\lambda_J^{2}$, and $\psi_{JI}$ are the functions of
$\lambda_I^{1}$, $\lambda_I^{2}$, $\psi_{IJ}$, and $r_{IJ}$:
\begin{eqnarray}
&&\lambda_J^{1} = \lambda_J^{1}\left(\lambda_I^{1}, \lambda_I^{2}, \psi_{IJ}, r_{IJ}\right)\label{eq;lambdaAlambdaI1}\;,\;\;\;\\
&&\lambda_J^{2} = \lambda_J^{2}\left(\lambda_I^{1}, \lambda_I^{2}, \psi_{IJ}, r_{IJ}\right)\label{eq;lambdaAlambdaI2}\;,\;\;\;\\
&&\psi_{JI} = \psi_{JI}\left(\lambda_I^{1}, \lambda_I^{2}, \psi_{IJ}, r_{IJ}\right)\label{eq;lambdaAlambdaI3}\;.\;\;\;
\end{eqnarray}
In (\ref{eq;lambdaAlambdaI1},\ref{eq;lambdaAlambdaI2},\ref{eq;lambdaAlambdaI3}) 
angles $\psi_{IJ}$ and $\psi_{JI}$ are the angles between the larger ellipses' axes and the direction
$\bm{r}_{IJ}$. Note that in isotropic elastic medium $\lambda_J^{1}$, $\lambda_J^{2}$, and $\psi_{JI}$ should 
not depend on the direction of $\bm{r}_{IJ}$ 
(the direction of the inclusion's deformation with respect to $\bm{r}_{IJ}$ is taken into account by the angle
$\psi_{IJ}$).

Note that the properties of the Eshelby's solution are embedded into
(\ref{eq;lambdaAlambdaI1},\ref{eq;lambdaAlambdaI2},\ref{eq;lambdaAlambdaI3}). 
These functions, in our view, represent the essence of the Eshelby's solution.
In Appendix \ref{app:inclusion} a particular case of the inclusion's shear transformation
is discussed and functions (\ref{eq;lambdaAlambdaI1},\ref{eq;lambdaAlambdaI2},\ref{eq;lambdaAlambdaI3})
for this case are derived.

Expressions for the stress correlation functions between the inclusion and the matrix
can be derived in the same way as the expressions 
(\ref{eq:sixysjxyref02},\ref{eq;F1},\ref{eq;F2},\ref{eq;F3},\ref{eq;pxy1},\ref{eq:sxysdiffcorrf01})
for the atomic level stress correlation functions.
For the product $\sigma^{xy}(I)\sigma^{xy}(J)$, for example, we get:
\begin{eqnarray}
&&\sigma^{xy}(I)\sigma^{xy}(J)\label{eq;ssxy2е}\\
&& = (1/8) \left[\mathcal{F}_1^e 
- \mathcal{F}_2^e\cos(4\theta_{IJ}) + \mathcal{F}_3^e\sin(4\theta_{IJ}) \right],\nonumber
\end{eqnarray} 
where
\begin{eqnarray}
&&\mathcal{F}_1^e \equiv (\lambda^1_I - \lambda^2_I)(\lambda^1_J - 
\lambda^2_J) \cos(2\psi_{IJ}-2\psi_{JI})\;,\;\;\;\label{eq;calF1e}\\
&&\mathcal{F}_2^e \equiv (\lambda^1_I - \lambda^2_I)(\lambda^1_J - 
\lambda^2_J) \cos(2\psi_{IJ}+2\psi_{JI})\;,\;\;\;\label{eq;calF2e}\\
&&\mathcal{F}_3^e \equiv (\lambda^1_I - \lambda^2_I)(\lambda^1_J - 
\lambda^2_J) \sin(2\psi_{IJ}+2\psi_{JI})\;.\;\;\;\label{eq;calF3e} 
\end{eqnarray}
The upper index $e$ in the formulas above originates from the word \emph{``elastic"}.
Note again that 
$\lambda_J^1$, $\lambda_J^2$, and $\psi_{JI}$ in 
(\ref{eq;calF1e},\ref{eq;calF2e},\ref{eq;calF3e}) are the functions of 
$\lambda_I^1$, $\lambda_I^2$, $\psi_{IJ}$, and $r_{IJ}$.
Also note that $\mathcal{F}_1^e$, $\mathcal{F}_2^e$, $\mathcal{F}_3^e$
do not depend on $\theta_{IJ}$.
Thus in (\ref{eq;calF1e},\ref{eq;calF2e},\ref{eq;calF3e})
\begin{eqnarray}
\mathcal{F}_n^e =\mathcal{F}_n^e(\lambda^1_I,\lambda^2_I,\psi_I,r_{IA})\;,\;\;\;\;\text{where}\;\;\;\;n=1,2,3\;,\;\;\;\;\label{eq;Fne}
\end{eqnarray}
i.e., functions
$\mathcal{F}_1^e$, $\mathcal{F}_2^e$, and $\mathcal{F}_3^e$ are determined
by how the stress field in the inclusion determines the stress field at point $J$.

Now we comment on the connection between the 
functions $\mathcal{F}_1^e$, $\mathcal{F}_2^e$, and $\mathcal{F}_3^e$
from (\ref{eq;calF1e},\ref{eq;calF2e},\ref{eq;calF3e}) and the functions
$F_1$, $F_2$, and $F_3$ from (\ref{eq;F1},\ref{eq;F2},\ref{eq;F3}).
The functions $\mathcal{F}_1^e$, $\mathcal{F}_2^e$, and $\mathcal{F}_3^e$
are written for a particular set 
of values $\lambda^1_I$, $\lambda^2_I$, $\psi_{IJ}$, and $r_{IJ}$.
In order to draw a parallel with the atomic level stress correlation functions
in liquids it is necessary to average the functions 
$\mathcal{F}_1^e$, $\mathcal{F}_2^e$, and $\mathcal{F}_3^e$ over
the possible values of $\lambda^1_I$, $\lambda^2_I$, and $\psi_{IJ}$ 
which can be associated with the parameters of the inclusion's 
stress ellipse. Thus:
\begin{eqnarray}
F_n^e(r_{IJ}) =\langle \mathcal{F}_n^e(\lambda^1_I,\lambda^2_I,\psi_I,r_{IJ})\rangle _{\lambda^1_I,\lambda^2_I,\psi_{IJ}}\;,\;\;\;\;\;\label{eq;FcalF1}
\end{eqnarray} 
where $n=1,2,3$. 
In (\ref{eq;FcalF1}) it is presumed that every set of parameters at $I$ deterministically
leads to certain parameters at $J$ via the Eshelby's solution.
In (\ref{eq;FcalF1}) there is no averaging over the distance (scalar) $r_{IJ}$.
Correspondingly functions $F_n^e$ depend only on $r_{IJ} \equiv r$.

In liquids there is no deterministic relation between the 
parameters and orientations of the atomic level stress ellipses of atoms $i$ and $j$.
In liquids there is only a probabilistic relation.
Thus in calculations of $F_1(r)$, $F_2(r)$, and $F_5(r)$ in liquids 
(\ref{eq;F1},\ref{eq;F2},\ref{eq;F3}) the averaging goes not only 
over $\lambda^1_i$, $\lambda^2_i$, $\psi_{ij}$, but also 
over $\lambda^1_j$, $\lambda^2_j$, $\psi_{ji}$. 
Implicitly in calculations of (\ref{eq;F1},\ref{eq;F2},\ref{eq;F3}) there is also
the averaging over the directions of $\bm{r}_{ij}$ for a fixed value of $r_{ij}$.
Since it is assumed that the undistorted inclusion and the matrix are 
isotropic there is no need to average (\ref{eq;FcalF1}) over the directions of $\bm{r}_{IJ}$.

Note that if $\langle \sigma_I^{xy}\sigma_J^{xy}\rangle $ were 
calculated from (\ref{eq;ssxy2}) in a particular reference frame,
by averaging over the possible distortions of the inclusion, it still would depend on $\theta_{IJ}$.
This dependence, however, would not reflect the essence of the angular dependent Eshelby's stress field. 
The dependence on $\theta_{IJ}$ in (\ref{eq;ssxy2}) reflects 
the rotational properties of the stress tensor.
The angular dependencies observed in Ref.\cite{Bin20151} correspond to the dependence
of $\langle \sigma^{xy}(I)\sigma^{xy}(J)\rangle $ on $\theta_{IJ}$ in (\ref{eq;ssxy2}). 
This is not the angular dependence of the Eshelby's field.
The angular dependence of the Eshelby's stress field is embedded
in how $\lambda^1_J$, $\lambda^2_J$, $\psi_{JI}$ depend on 
$\lambda^1_I$, $\lambda^2_I$, $\psi_{IJ}$, and $r_{IJ}$.

\begin{figure*}
\begin{center}
\includegraphics[angle=0,width=6.0in]{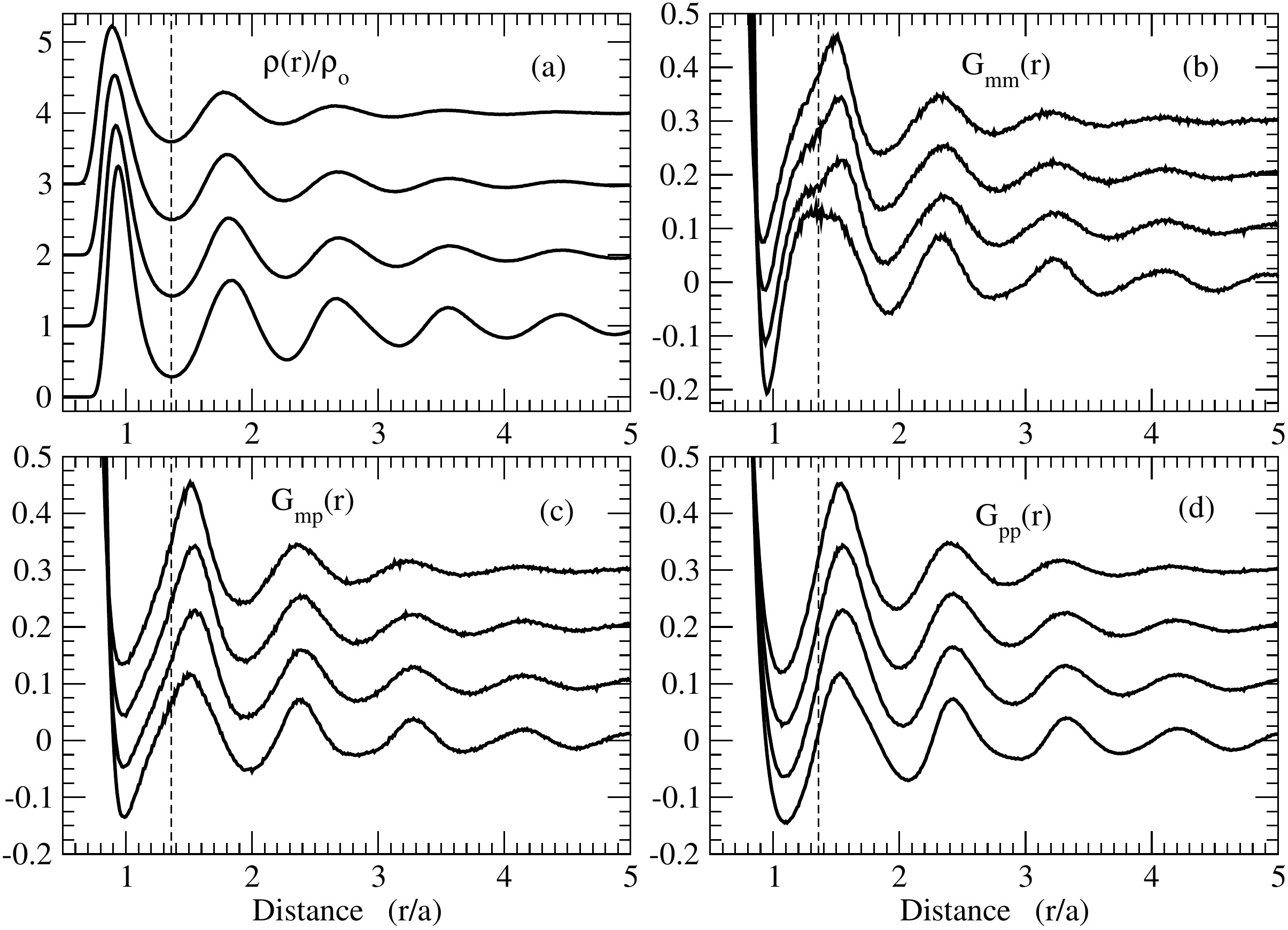}
\caption{In every panel the curves from the top to the bottom correspond to temperatures:
$T=3$, $T=2$, $T=1.4$, and $T=1$. 
The curves were shifted vertically for the clarity of the presentation.
{\bf(a)} Pair density function. 
{\bf (b)} $G_{mm}$ correlation function from (\ref{eq;Gmm}).
{\bf (c)} $G_{mp}$ correlation function from (\ref{eq;Gmp}).
{\bf (d)} $G_{pp}$ correlation function from (\ref{eq;Gpp}).
}\label{fig:rhor-allG}
\end{center}
\end{figure*}

\begin{figure}
\begin{center}
\includegraphics[angle=0,width=2.8in]{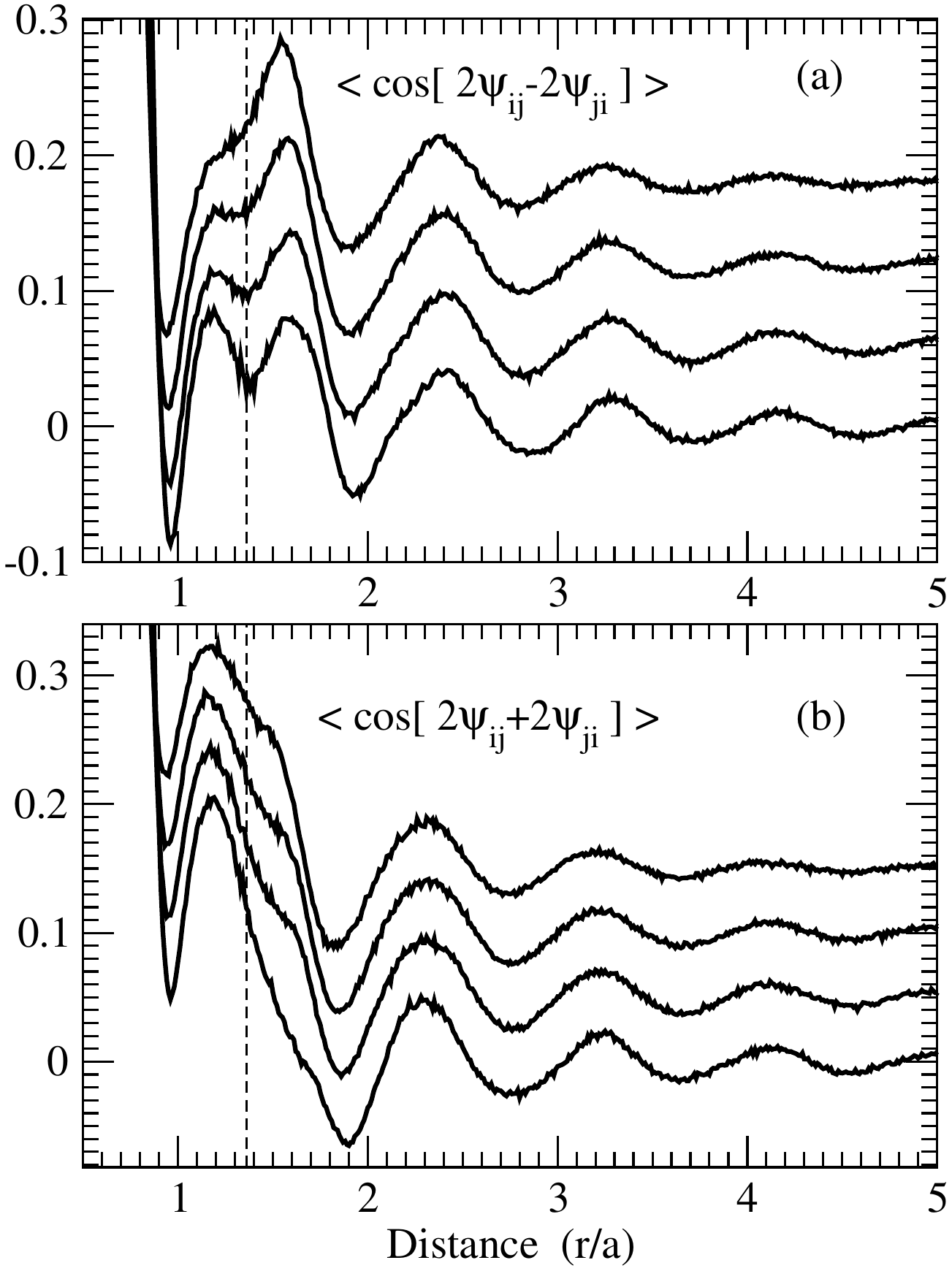}
\caption{Evolutions with temperature of the 
functions $\langle \cos(2\psi_{ij}-2\psi_{ji})\rangle $ 
and $\langle \cos(2\psi_{ij}+2\psi_{ji})\rangle $. 
In both panels the curves from the top to the bottom correspond 
to temperatures $T=3$, $T=2$, $T=1.4$, and $T=1$. 
The curves were shifted vertically for the clarity of the presentation.
}\label{fig:cosMM-1}
\end{center}
\end{figure}

\begin{figure}
\begin{center}
\includegraphics[angle=0,width=3.3in]{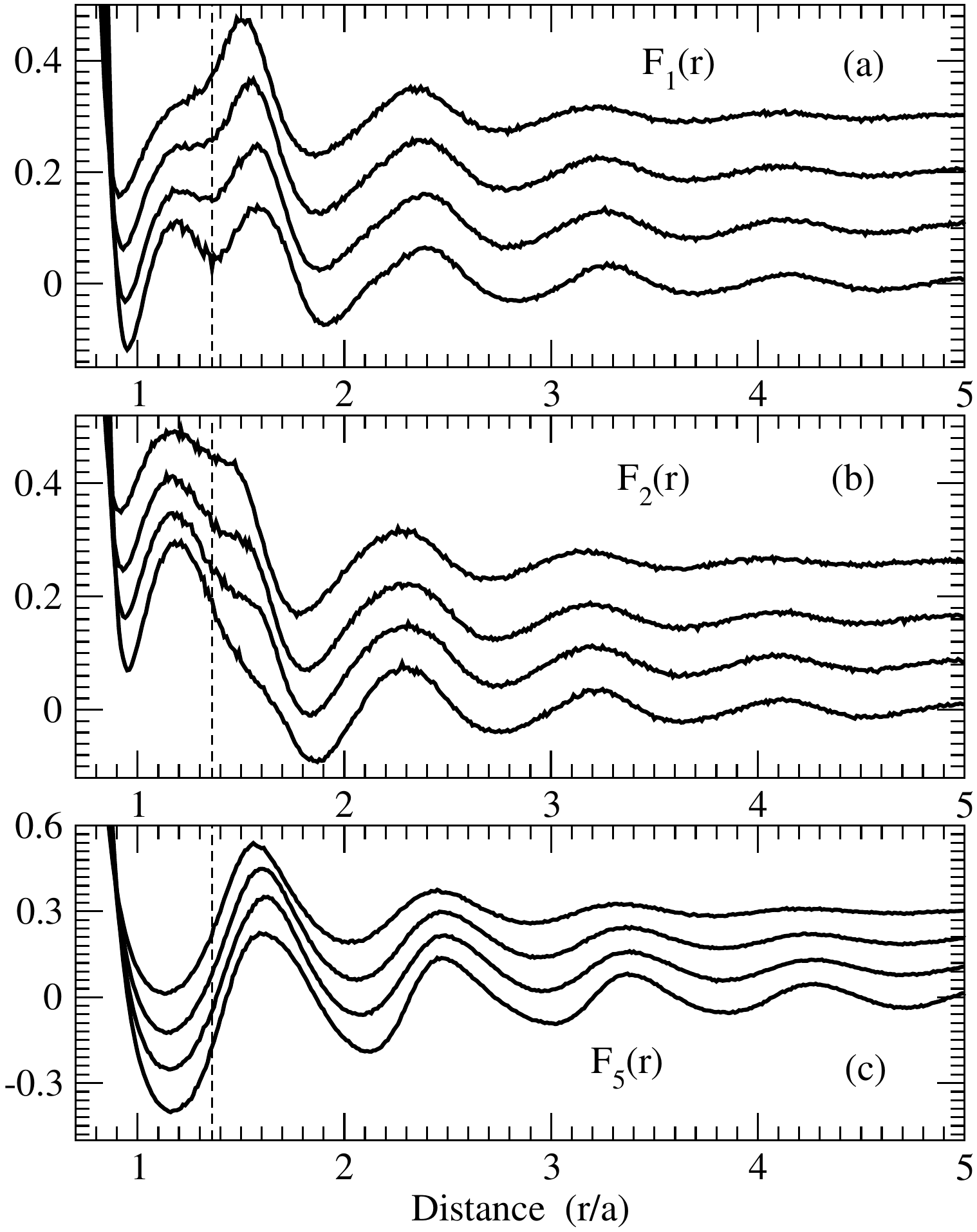}
\caption{
Evolutions with temperature of the {\it normalized} functions $\tilde{F}_1$ 
from (\ref{eq;F1},\ref{eq;tF1}), 
$\tilde{F}_2$ from (\ref{eq;F2},\ref{eq;tF1}), 
and $\tilde{F}_5$ from (\ref{eq;F5},\ref{eq;tF2}).
In every panel the curves from the top to the bottom correspond 
to temperatures $T=3$, $T=2$, $T=1.4$, and $T=1$. 
The curves were shifted vertically for the clarity of the presentation.
} 
\label{fig:F1F2F5-1}
\end{center}
\end{figure}

\begin{figure}
\begin{center}
\includegraphics[angle=0,width=3.3in]{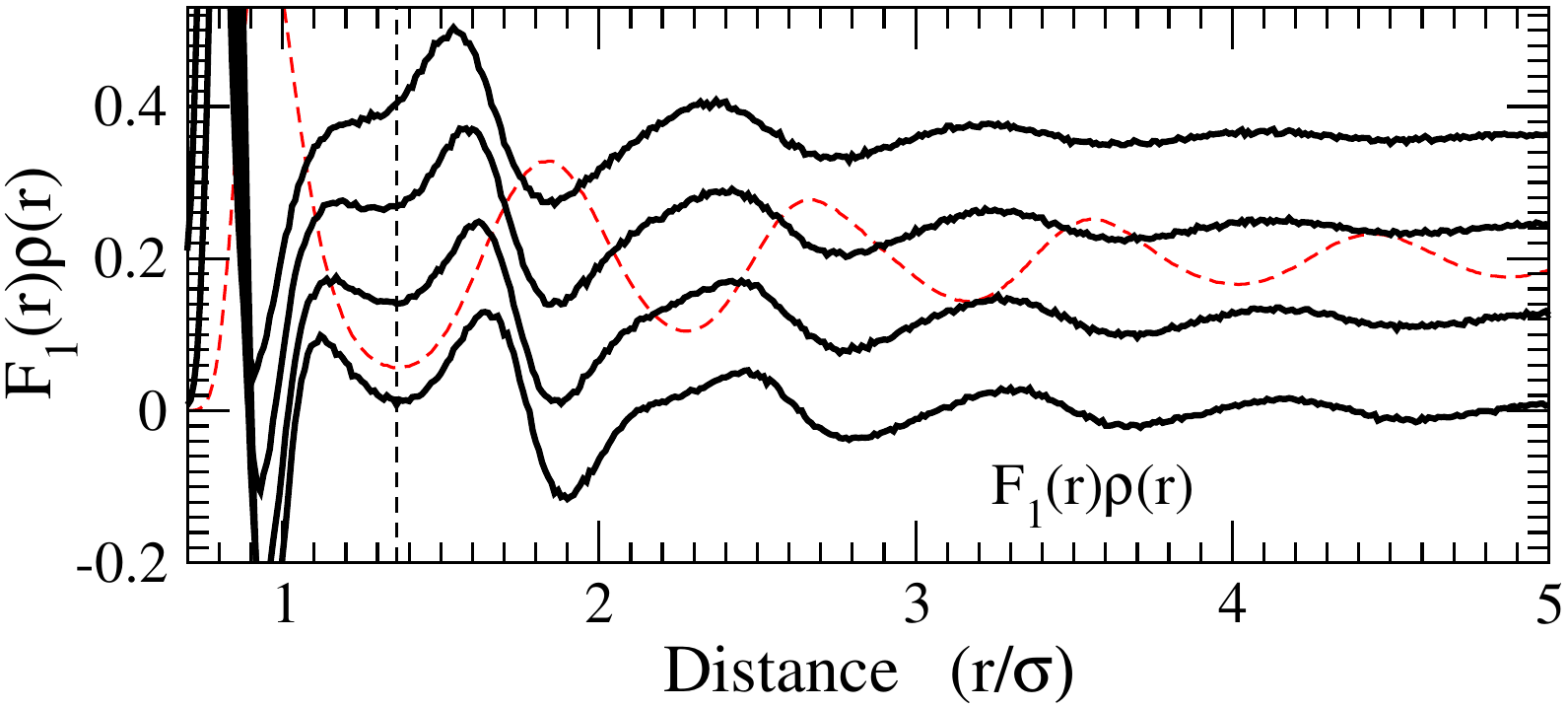}
\caption{
Evolution with temperature of the function $\tilde{F}_1(r)\rho(r)$.
The curves from the top to the bottom correspond 
to temperatures $T=3$, $T=2$, $T=1.4$, and $T=1$. 
The curves were shifted vertically for clarity of presentation.
The dashed curve shows the scaled $\rho(r)$ at $T=1.0$.
As temperature decreases the first minimum in $\rho(r)$
becomes deeper. This deepening overlaps with the development
of the minimum in $\tilde{F}_1(r)$. 
Thus, the development of the feature in $\tilde{F}_1(r)$ at the position 
of the first minimum of $\rho(r)$ is also well pronounced in $\tilde{F}_1(r)\rho(r)$ .
} 
\label{fig:F1rhor-1}
\end{center}
\end{figure}

\begin{figure*}
\begin{center}
\includegraphics[angle=0,width=6.0in]{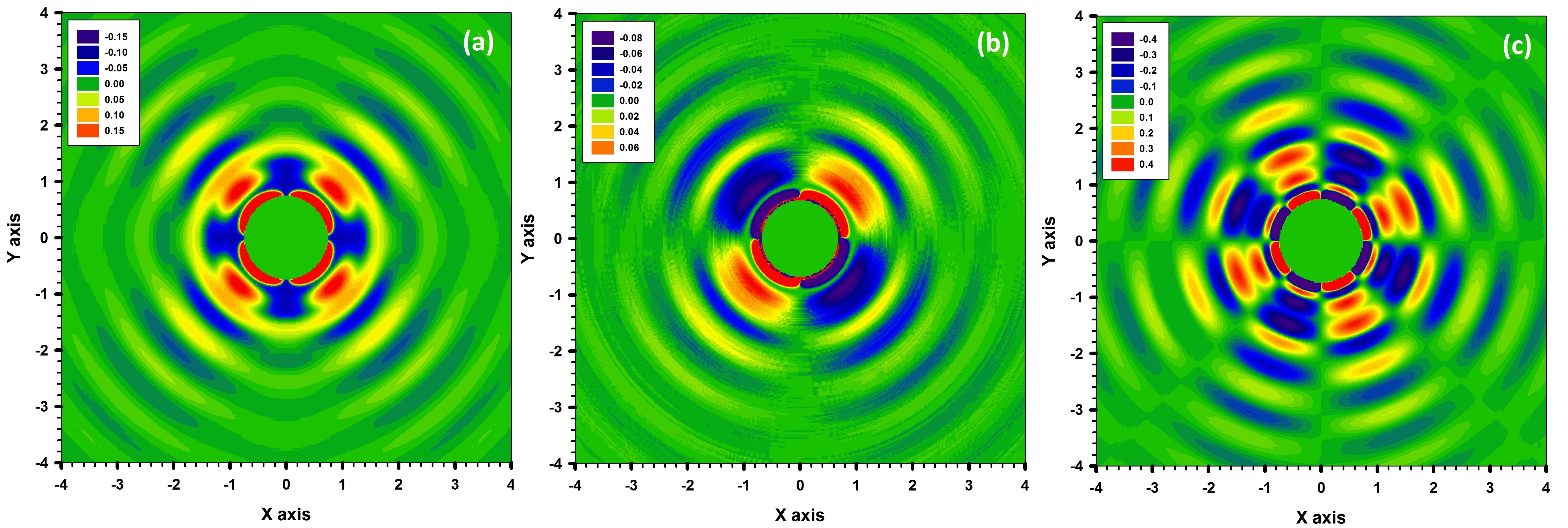}
\caption{{\bf (a)} Atomic level stress correlation function 
$\langle \sigma_i^{xy}\sigma_j^{xy}\rangle $.
See formula (\ref{eq:sixysjxyref02}). 
The functions $F_1$ and $F_2$ 
in $\langle \sigma_i^{xy}\sigma_j^{xy}\rangle $ 
were normalized according to (\ref{eq;tF1}). 
The function $F_3$ is zero, besides the noise.
It is clear that panel (a) of this figure is very similar to the 
panel (a) of Fig.5 in Ref.\cite{Bin20151}.
{\bf (b)} Atomic level stress correlation function
$\langle p_i\sigma_j^{xy}\rangle $. 
See formula (\ref{eq;pxy1}). 
The function $F_4$ in $\langle p_i\sigma_j^{xy}\rangle $ averages to zero.
Thus only function $F_5$ is left. In order to produce this figure
we had to subtract the average pressure from the diagonal components
of the atomic level stress tensor. Effectively this means that 
the value of $\lambda$ averaged over $\lambda^1$ and $\lambda^2$ of 
all atoms, i.e., $\lambda^{ave}$, was subtracted from the values
of $\lambda^1$ and $\lambda^2$ of every atom. 
Because of this subtraction we can not use normalization 
(\ref{eq;Gmp}) since $\langle \lambda^1 + \lambda^2 - 2\lambda^{ave}\rangle $ 
averages to zero. Thus we used normalization (\ref{eq;Gmm}) instead. 
We also scaled intensity on the $z$-axis by a factor of 4.
It is obvious that panel (b) of this figure is very similar to the panel
(b) of Fig.5 in Ref.\cite{Bin20151}.
{\bf (c)} Atomic level stress correlation function
$\langle (\sigma^{xx}_i-\sigma^{yy}_i)\sigma_j^{xy}\rangle $.
See Eq.(\ref{eq:sxysdiffcorrf01}). 
Only function $F_1$ in 
$\langle (\sigma^{xx}_i-\sigma^{yy}_i)\sigma_j^{xy}\rangle $ 
is non-zero.
We scaled the function by a factor of $10$ along the $z$-intensity axis.
It is clear that correlation function in panel
(c) is rather similar to the correlation function in panel
(c) of Fig.5 in Ref.\cite{Bin20151}. There is a difference with the two
``circles" at $(r/a) \lesssim  1$. It is possible that authors 
of Ref.\cite{Bin20151} did not look into such small distances and thus 
in their figure these ``circles" fall into the central green region.
Besides this difference their figures and ours look rather similar.
}\label{fig:2Dplots}
\end{center}
\end{figure*}

\section{Results of MD simulation \label{sec:MDresults}}

\subsection{Stress correlation functions}

In our Molecular Dynamics (MD) simulations we considered the same 2D system of particles
that has been studied in Ref.\cite{Bin20151}. We used the same Yukawa potential and 
the same LAMMPS MD program  \cite{Plimpton1995,lammps}.
We studied the systems of two sizes.
In the \emph{small} system the number of particles was $N=2500$, while
the dimensions of the rectangular periodic box 
were $L_x=50.1021$, $L_y=43.3897$. 
Our \emph{small} system has the same size as the system studied in Ref.\cite{Bin20151}.
Another \emph{(large)} system that we studied contained $N=22500$ particles, 
i.e., nine times more than the \emph{small} system.
The dimensions of the \emph{large} system were $L_x = 150.306$, $L_y=130.169$.
The particles' number densities in the \emph{small} and \emph{large} systems
are the same.
We performed simulations in NVT and NVE ensembles.

In all cases the systems were prepared by 
melting triangular lattice at reduced temperature $T=5$. After 
the equilibration at $T=5$ the temperature was reduced in several 
steps that followed by equilibration at every temperature (in NVT ensemble) 
or at every value of fixed total energy (in NVE ensemble).
The temperature in NVT ensemble was introduced via Nos\'{e}-Hoover thermostat.
The damping parameter corresponded to 100 MD steps and also to 0.1 of the time unit.

In our simulations, we reproduced the 
dependence of potential energy on temperature presented in Fig.~1 of 
Ref.~\cite{Bin20151}.

Atomic configurations for calculations of the correlation functions 
related to the eigenvalues and eigenvectors
of atomic level stresses were 
collected on the \emph{small} system in the NVE ensemble at total energies 
which corresponded to the following temperatures: 
$T(3)=3.06 \pm 0.04$, $T(2)=1.97 \pm 0.03$, $T(1.4)=1.43 
\pm 0.02$, $T(1)=0.99 \pm 0.02$. 
The averaging was done over 1000 configurations at every temperature.
For the temperature $T=1$ the time interval between the two 
consecutive configurations was $10^4$ MD steps.
Each MD step corresponded to 0.001 of the time unit. 
During these $10^4$ MD steps the 
mean square atomic displacement reaches $\sim 1.38\sigma$. 

Different correlation functions {\it per 
pair of particles} are shown in 
Figs.\ref{fig:rhor-allG},\ref{fig:cosMM-1},\ref{fig:F1F2F5-1}. The 
dependencies of the functions $\tilde{F}_1$, $\tilde{F}_2$ and 
$\tilde{F}_5$, i.e., all the non-zero ones, on distance are shown in 
Fig.~\ref{fig:F1F2F5-1}. At $T = 1$ we have $Z_{-} = 
\langle  \lambda_i^1 - \lambda_i^2 \rangle \approx 10.82$ and $Z_{+} = 
\langle  \lambda_i^1 + \lambda_i^2 \rangle \approx 40.48$.

Figures \ref{fig:rhor-allG},\ref{fig:cosMM-1},\ref{fig:F1F2F5-1} 
demonstrate that there are 
$r_{ij}$-dependent correlations between the parameters of the atomic 
level stress ellipses and in their orientations. These correlations 
gradually decrease with increase of $r_{ij}$. 
It is clear that functions $G_{mm}(r)$ in Fig.\ref{fig:rhor-allG}(b) 
and $<\cos(2\psi_{ij}-2\psi_{ji})>$ in Fig.\ref{fig:cosMM-1}(a)
exhibit more pronounced changes than does PDF [Fig.\ref{fig:rhor-allG}(a)] 
on decrease of temperature.
It is also clear that the first peaks in
$\tilde{F}_1$ and $\tilde{F}_2$ [Fig.\ref{fig:F1F2F5-1}(a,b)] 
also demonstrate more pronounced changes on decrease of temperature than does PDF.
However, it is also more difficult to interpret these changes. Yet, developing
features in $\langle\cos(2\psi_{ij} - 2\psi_{ji})\rangle$ suggest that some ordering happens
in the mutual orientations of the ellipses associated with the atoms separated by the distance 
corresponding the first minimum in the PDF. 
There also appears to be a certain similarity in the behaviours of
$\langle \cos(2\psi_{ij}-2\psi_{ji})\rangle$ and $\tilde{F}_1$. 
This similarity suggests that changes in $\tilde{F}_1$ are caused by 
changes in $\langle \cos(2\psi_{ij}-2\psi_{ji})\rangle$. 
See expression (\ref{eq;F1}) for $F_1$. 
Thus changes in $\tilde{F}_1$ are likely to be caused
not by changes in the eigenvalues of the stress ellipses, 
but by changes in the mutual orientations of the ellipses. 
However, also note that there are changes in
$G_{mm}(r)$ in Fig.\ref{fig:rhor-allG}(b).

Figure (\ref{fig:F1rhor-1}) shows how the 
function $\tilde{F}_1(r)\rho(r)$ changes with temperature.
It follows from the figure that as temperature is reduced there 
develops a pronounced minimum at the position of the first minimum, $r^1_{min}$,
of $\rho(r)$. Thus changes in $\tilde{F}_1(r)$ are also well observable in
$\tilde{F}_1(r)\rho(r)$ despite the fact that
the number of atomic pairs separated by  $r^1_{min}$ is relatively small.

The curves in Fig.\ref{fig:F1F2F5-1} can be converted 
into the 2D intensity plots equivalent to those presented in Ref.\cite{Bin20151}
using formulas (\ref{eq:sixysjxyref02},\ref{eq;pxy1},\ref{eq:sxysdiffcorrf01}). 
Thus, if we want to find the stress correlation function at a point
with coordinates $(x,y)$ we define $r=\sqrt{x^2+y^2}$ and 
$\theta=\arctan{(y/x)}$. 
Using these values in (\ref{eq:sixysjxyref02},\ref{eq;pxy1},\ref{eq:sxysdiffcorrf01}) 
the stress field at $(x,y)$ can be found.
This conversion applies because for particles $i$ and $j$ 
with coordinates $(x_i,y_i)$ and $(x_j,y_j)$ 
the values of $r_{ij}$ and $\theta_{ij}$ that go into the formulas
(\ref{eq:sixysjxyref02},\ref{eq;pxy1},\ref{eq:sxysdiffcorrf01}) are
$r_{ij}=\sqrt{(x_j-x_i)^2+(y_j-y_i)^2}$ 
and $\theta_{ij}= \arctan{((y_j-y_i)/(x_j-x_i))}$.
However, in making the $2D$ stress correlation function plots it is assumed that
the particle $i$ is at the origin.

The results of the conversion described above for $T=1$ are presented in
Fig.\ref{fig:2Dplots}.
It is obvious that the 2D plots in Fig.\ref{fig:2Dplots}. 
are very similar to those shown in Fig.5 of Ref.\cite{Bin20151}. 
Note that the $2D$ plots presented in Fig.\ref{fig:2Dplots} were obtained
from only 3 functions, i.e., $F_1(r)$, $F_2(r)$, and $F_5(r)$ which 
depend only on $r$. 
This proves that the dependencies on $\theta$ presented in the 2D plots in
Ref.\cite{Bin20151} follow from the tensorial rotational properties.

\subsection{Is system studied a true liquid or is it in a hexatic phase?}

Finally, we comment on the following statement made in Ref.\cite{Bin20151}.
It is stated there that at $T=1$ the system is in a true liquid state, while
at $T=0.95$ the system is in a hexatic state.

In order to make a distinction between the true liquid and haxatic states
it was assumed in Ref.\cite{Bin20151} that in a true liquid state bond-order
correlation function decays exponentially with increase of distance, while
in the haxatic state the bond-order correlation function decays algebraically.

\begin{figure}
\begin{center}
\includegraphics[angle=0,width=3.3in]{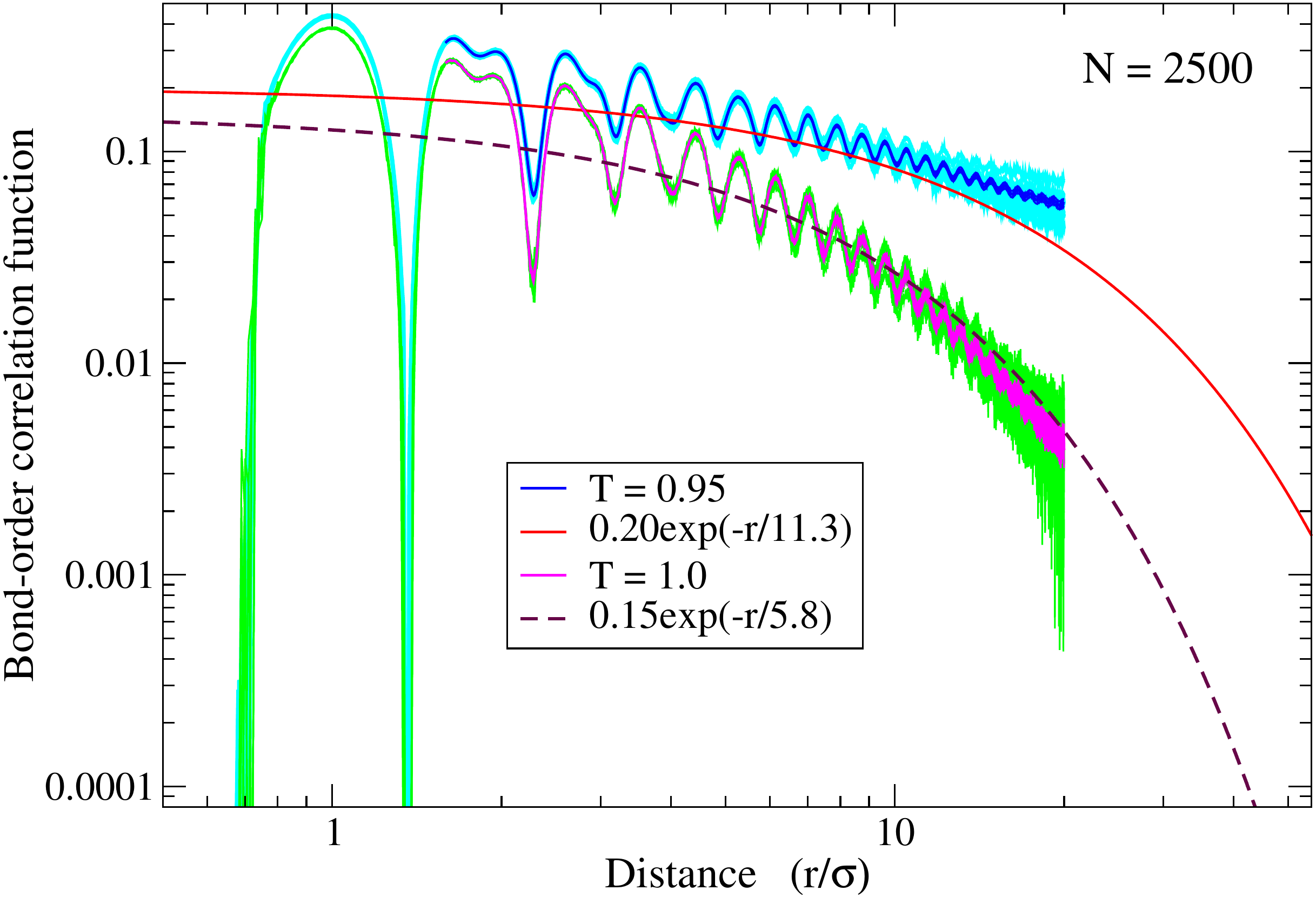}
\caption{Bond-order correlation functions for the system of $N=2500$ particles.
There are 10 cyan curves in the figure. Every cyan curve is the average over
1000 independent configurations. The mean square displacement between the consecutive
configurations was larger than the interatomic distance. 
There are three blue curves in the figure.
They represent the mean over the 10 cyan curves and the average $\pm$ the error of the mean.
There are 10 green curves in the figure. 
Every green curve represents the average over 100 independent configurations.
There are 3 magenta curves which represent the average over the green curves 
and the average $\pm$ the error of the mean. 
The red and maroon curves show the fits obtained
from the larger system of $N=22500$.
}\label{fig:bocf2500}
\end{center}
\end{figure}

\begin{figure}
\begin{center}
\includegraphics[angle=0,width=3.3in]{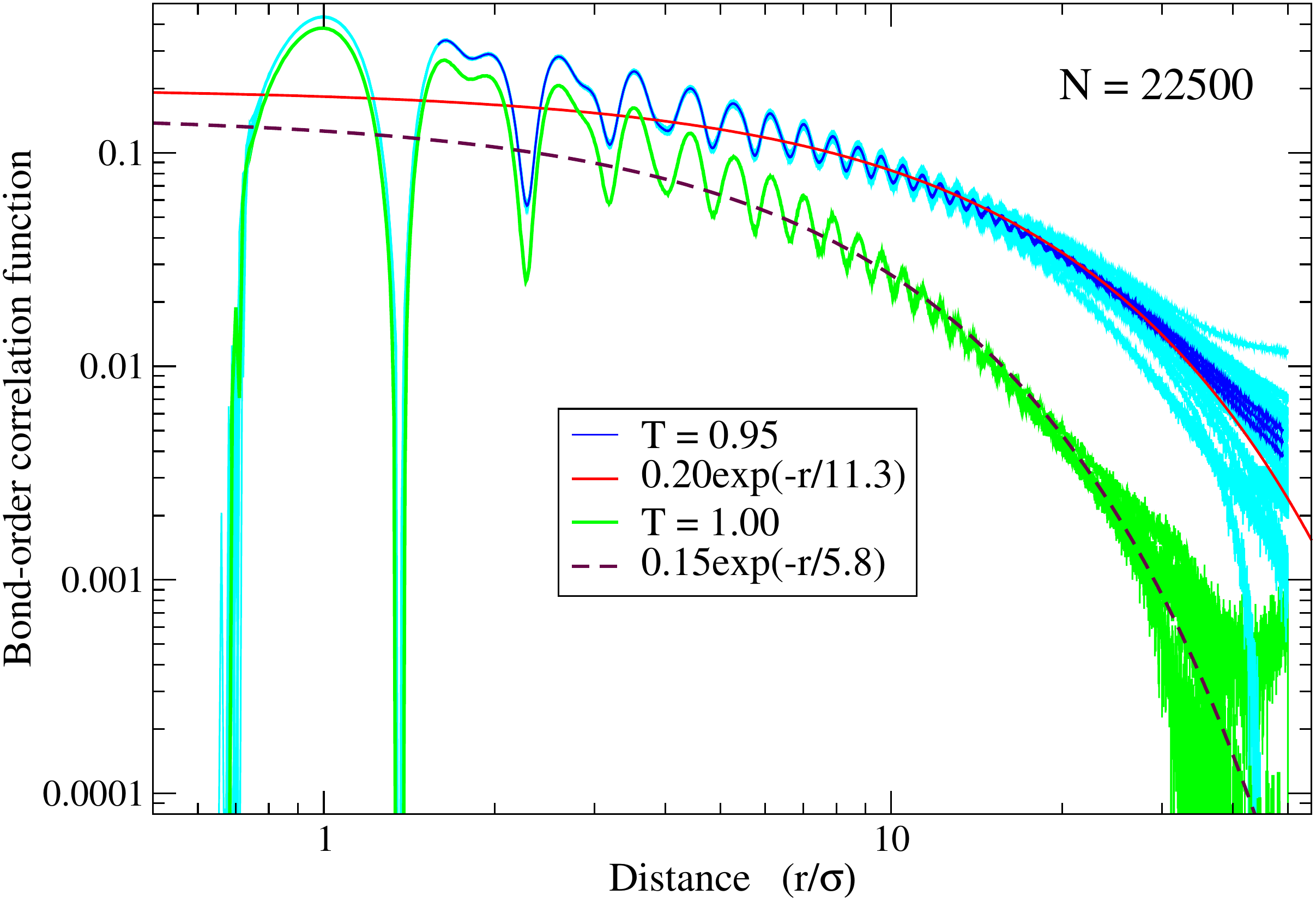}
\caption{Bond-order correlation functions for the system of $N=22500$ particles.
There are 21 cyan curves in the figure. Every cyan curve is the average over
100 independent configurations. The mean square displacement between the consecutive
configurations was larger than the interatomic distance. 
There are three blue curves in the figure.
These blue curves represent the mean over the 21 cyan curves and the 
mean $\pm$ the error of the mean curves.
It is possible that at distances larger than $r/\sigma=30$ at $T=0.95$ 
there again appear size effects.
There are three green curves in the figure. 
Every green curve is the average over 200 independent configurations.
The red and maroon curves are the fits to the data.
}\label{fig:bocf22500}
\end{center}
\end{figure}

We calculated how the bond-order correlation function depends on distance in 
systems of two sizes. In the small system containing $N=2500$ particles
$(L_y/2) = 21.7$, while $(L_y/2) = 65.1$ in the large system with $N=22500$.
The small system was used in Ref.\cite{Bin20151}.
In our calculations we assumed that two atoms are the nearest neighbours 
if they are separated by a distance smaller than the position of the 
first minimum in the PDF, i.e., $(r_{ij}/a) \leq 1.36\sigma$.
The results are presented in Fig.\ref{fig:bocf2500},\ref{fig:bocf22500}.

It follows from Fig.\ref{fig:bocf2500},\ref{fig:bocf22500} that on decrease of temperature
haxatic order undoubtedly develops in the systems.
The comparison of Fig.\ref{fig:bocf2500} with Fig.\ref{fig:bocf22500} 
suggests that at $T=0.95$ the small system exhibits observable size effects. 
Note that at $T=0.95$ the decay length is larger than (1/2) 
of $L_y/2$ in the small system. 
Thus in the small system the bond-order correlation function
does not decay completely on the length of the half of the simulation box.
It also follows from the data obtained on the large system that
exponential fit to the data is better than can be any algebraic fit at both temperatures. 
Thus, in our view, it follows from Fig. \ref{fig:bocf2500},\ref{fig:bocf22500}
that it is impossible to make a qualitative distinction between the liquid
states at $T=1$ and $T=0.95$. The observation of the algebraic decay
at $T=0.95$ reported in Ref.\cite{Bin20151} is probably related to the size effects.

\section{Conclusion \label{sec:concl}}

It was demonstrated that it is possible to study liquid (and glass) 
structures through considerations of correlations between the 
eigenvalues and eigenvectors of the atomic level stress tensors of 
different atoms. It was shown that on decrease of temperature some of 
the studied correlation functions exhibit pronounced changes in the range
of distances that corresponds to the first minimum of the pair density function.
These changes could not be guessed from the behaviour of the pair density function. 
Thus the suggested method provides additional information and it is of interest to 
investigate evolution of stress correlations with this method in model 
supercooled liquids on decrease of temperature.

We also demonstrated that interpretations of the angular dependencies of the stress 
correlation functions reported in 
Ref.\cite{Bin20151} are essentially incorrect. 
In particular, the authors of Ref.\cite{Bin20151} associate
the angular dependencies observed in the stress correlation functions with 
the angular dependencies of the Eshelby's stress field.
We demonstrated that anisotropic stress fields observed in Ref.\cite{Bin20151}
originate from the rotational properties of the stress tensors.
We also had shown that information which is really related to the anisotropic Eshelby's stress 
fields is embedded into the isotropic stress correlation functions $F_1(r)$, $F_2(r)$, and $F_5(r)$
which we studied in this work.  

From a purely pragmatic perspective we have shown that eight 2D-panels
of the stress correlation functions presented in Ref.\cite{Bin20151} can be reproduced using
only 3 correlation functions which depend only on $r$, i.e.,
from the function $F_1(r)$, $F_2(r)$, and $F_5(r)$ \cite{2Dplots1}. 
This clearly advances understanding of the plots of the stress correlation functions
presented in Ref.\cite{Bin20151}. 
It also follows from our results that instead of studying distance dependence 
of the integrals of the 2D stress correlation functions over some angles, 
as it has been done in Ref.\cite{Bin20151}, it is more 
reasonable to study how functions $F_1(r)$, $F_2(r)$, and $F_5(r)$ 
depend on distance.

We also demonstrated that because of size effects the distinction
made in Ref.\cite{Bin20151} between the normal liquid and 
haxatic states is invalid.

\appendix

\section{Alternative derivation of $\langle  \sigma_i^{xy} \sigma_j^{xy} 
\rangle$ structure\label{sec:mishapdx}}

In 2D in a particular reference coordinate frame numerical 
representation of the atomic level stress tensor ${\hat\sigma}$ 
is a $2 \times 2$ matrix. This matrix is real and symmetric ({\it i.e.}, 
$\sigma^{yx} = \sigma^{xy}$), thus it can be diagonalized. 
We can work directly with its components $\sigma^{\alpha\beta}$; or with 
corresponding pressure $p$ and two shear components, $s_1$ and $s_2$; 
or with real eigenvalues $\lambda_{1,2}$ and the 
orientation of two orthogonal eigenvectors:
 \begin{align*}
   {\hat\sigma} &= \left[ \begin{array}{cc}
     \sigma^{xx} & \sigma^{xy} \\
     \sigma^{xy} & \sigma^{yy}
   \end{array} \right] =
   \left[ \begin{array}{cc}
     p + s_1 & s_2 \\
     s_2 & p - s_1
   \end{array} \right] \\
   &= {\hat{R}}(\varphi) \left[ \begin{array}{cc}
     \lambda_1 & 0 \\
     0 & \lambda_2
   \end{array} \right] \bigl( {\hat{R}}(\varphi) \bigr)^T .
 \end{align*}
 Here ${\hat{R}}(\varphi)$ is the $2 \times 2$ matrix of rotation in 
positive (or counterclockwise) direction by angle $\varphi$:
 \begin{align*}
   {\hat{R}}(\varphi) &= \left[ \begin{array}{rr}
     \cos \varphi & ~-\sin \varphi \\
     \sin \varphi &   \cos \varphi
   \end{array} \right] .
 \end{align*}

Pressure and shear components are expressed through $\sigma^{\alpha\beta}$ components as
 \begin{align*}
   & p = {\textstyle\frac{1}{2}} \bigl( \sigma^{xx} + \sigma^{yy} \bigr) 
\,, \quad
   s_1 = {\textstyle\frac{1}{2}} \bigl( \sigma^{xx} - \sigma^{yy} \bigr) 
\,, \quad
   s_2 = \sigma^{xy} \,.
 \end{align*}

 It will be convenient for us to combine the shear components into a 
single complex number $s = s_1 + i \, s_2$. The total amount of shear is 
given by its absolute value:
 \begin{align*}
   |s| = \sqrt{s_1^2 + s_2^2} = \sqrt{ {\textstyle\frac{1}{4}} 
   (\sigma^{xx} - \sigma^{yy})^2 + (\sigma^{xy})^2 } \,,
 \end{align*}
 while the argument of $s$ is related to the shear's direction.
 
 Consider three reference frames $(x, y)$, $(x', y')$, and $(x'', y'')$, 
with $(x', y')$ being obtained from $(x, y)$ by a rotation in negative 
(clockwise) direction by angle $\alpha$, while $(x, y)$ and $(x'', y'')$ 
are mirror reflections of each other with respect to $x$-axis. We will 
write down the quantities in $(x', y')$ and $(x'', y'')$ frames with 
prime and double prime symbols, respectively. The transformation 
properties of the stress 
tensor (\ref{eq:rotABxx},\ref{eq:rotAByy},\ref{eq:rotABxy})
 result in
 \begin{align*}
   & s' = s \, \exp(2 i \, \alpha) \,, \quad s'' = s^* \,,
 \end{align*}
 where $\cdot^*$ denotes complex conjugation.

 Since $\sigma^{xy} = (s - s^*) / 2 i$, we have:
 \begin{align}
  \langle  \sigma_i^{xy} \sigma_j^{xy} \rangle &= 
{\textstyle\frac{1}{16}} \bigl[F_1(r, \theta) + F_1^*(r, \theta) - F_2(r, \theta) - F_2^*(r, \theta)\bigr] \,, 
\label{s2s2_FFFF1}
 \end{align}
where
 \begin{align}
   F_1(r, \theta) &= 4\langle  s_i s^*_j \rangle \,, \quad
   F_2(r, \theta) = 4\langle  s_i s_j \rangle \,;
\label{s2s2_FFFF2}   
 \end{align}

 All the averages $\langle  \cdot \rangle$ are taken over the pairs of 
atoms $i$ and $j$ with $r_{i\!j} = r$ and $\theta_{i\!j} = \theta$.

 By checking how $s$ and the angle $\theta$ are transformed by rotations 
($s' = s \, \exp(2 i \, \alpha)$ and $\theta' = \theta + \alpha$) we get
 \begin{align}
   F_1(r, \theta + \alpha) &= F_1(r, \theta) \,, \nonumber \\
   F_2(r, \theta + \alpha) &= F_2(r, \theta) \, \exp(4 i \, \alpha) \,. 
\label{transform_F_2}
 \end{align} The function $F_1(r, \theta)$ does not depend on the angle 
$\theta$ at all. By considering $F_1^*(r, \theta)$ we exchange the roles of atoms 
$i$ and $j$, which is equivalent to the change $\theta \to \theta + 
\pi$. Thus $F_1^*(r, \theta)=F_1(r, \theta + \pi) = F_1(r, \theta)$, i.e.,
we get $F_1 = F_1^*$. 
All this means that $F_1(r, \theta) = F_1(r)$ is a real function of a 
single parameter $r$.

 If we put $\theta = 0$ in \eqref{transform_F_2}, we get $F_2(r, \alpha) 
= F_2(r, 0) \, \exp(4 i \, \alpha)$. Mirror reflection ($s'' = s^*$ and 
$\theta'' = -\theta$) leads to $F_2(r, -\theta) = F_2^*(r, \theta)$. In 
particular, $F_2(r) = F_2(r, 0) = F_2^*(r, 0)$ is also a real function 
of just the distance between the atoms $r$. Also, $F_2(r, \theta) = 
F_2(r) \, \exp(4 i \, \theta)$.

Putting these results for $F_1$ and $F_2$ into the expression 
\eqref{s2s2_FFFF1} we finally get
 \begin{align}
   \langle  \sigma_i^{xy} \sigma_j^{xy} \rangle &= 
{\textstyle\frac{1}{8}} \bigl( F_1(r) - F_2(r) \cos(4 \theta) \bigr) \,. 
\label{s2s2_FF}
 \end{align}
 Note that the angular dependence of this correlation function was 
obtained solely by checking how the atomic level stress tensors are 
transformed under rotations (and mirror reflections). Thus the physical 
properties of the liquid prescribe the $r$-dependence of $F_{1,2}(r)$, 
but not the $\theta$-dependence in \eqref{s2s2_FF}.

\section{Yet, another derivation of the expression for $\langle  \sigma_i^{xy} 
\sigma_j^{xy} \rangle $ \label{sec:derivation3}}

Let us suppose that the first eigenvectors of atoms $i$ and $j$ form angles
$\varphi_i$ and $\varphi_j$ with the $\hat{x}$-axis of our reference coordinate frame.
See Fig.\ref{fig:angles}.

From (\ref{eq:rotixy}) the product $\sigma_i^{xy} \sigma_j^{xy}$ in our reference coordinate frame has the form:
 \begin{eqnarray}
   \sigma_i^{xy}\sigma_j^{xy}= (1/4)(\lambda^1_i - 
\lambda^2_i)(\lambda^1_j -\lambda^2_j)
   \sin(2\varphi_i)\sin(2\varphi_j)\label{eq;ssxy1}\;.\;\;\;\;\;\;\;\;
 \end{eqnarray}
Note that the dependence on angles $\varphi_i$ and $\varphi_j$ appears in
(\ref{eq;ssxy1}) from the ``rotations" (\ref{eq:rotixy}) of the stresses 
from the coordinate frames of their eigenvectors into our reference coordinate frame.
Thus dependence of (\ref{eq;ssxy1}) on $\varphi_i$ and $\varphi_j$ reflects
transformational properties of the stress tensors under rotations.
 
We then, using Fig.\ref{fig:angles}, express angles $\varphi_{i}$ and $\varphi_{j}$ through the 
angles $\psi_{ij}$, $\psi_{ji}$, and $\theta_{ij}$: 
$\varphi_i = \psi_{ij} + \theta_{ij}$ and $\varphi_j = \psi_{ji} + \theta_{ij}$.
Substitution of these expressions for $\varphi_i$ and $\varphi_j$ into (\ref{eq;ssxy1})
(with the following averaging) leads to:
 \begin{eqnarray}
   \langle \sigma_i^{xy}\sigma_j^{xy}\rangle _{\theta_{ij}=\theta} = (1/8) \left[F_1 
   - F_2\cos(4\theta) + F_3\sin(4\theta) \right] 
   \label{eq;ssxy2},\;\;\;\;\;\;\;
 \end{eqnarray} 
where $F_1$, $F_2$, and $F_3$ are given by expressions (\ref{eq;F1},\ref{eq;F2},\ref{eq;F3}).
Note that (\ref{eq;ssxy2}) is identical to (\ref{eq:sixysjxyref02}) and (\ref{s2s2_FF}).

Since the dependence of (\ref{eq;ssxy1}) on $\varphi_i$ and $\varphi_j$ appeared
from the rotational properties of the stress tensors the dependence of (\ref{eq;ssxy2})
on $\theta$ also reflects the rotational properties of the stress tensors.

\section{Eshelby's stress field in the directional frame for a case
of shear deformation of a circular inclusion. Functions
$\mathcal{F}_1^e$, $\mathcal{F}_2^e$, $\mathcal{F}_3^e$ \label{app:inclusion}}

In this section we derive the expressions relating
the eigenvalues and eigenvectors of the stress fields in the inclusion and in the matrix
for a particular case of unconstrained shear strain applied to the inclusion. 
Then we calculate functions 
$\mathcal{F}_1^e$, $\mathcal{F}_2^e$, $\mathcal{F}_3^e$
for the considered example.
We start from the known formulas for the Eshelby's stress field
\cite{Eshelby1957,Eshelby1959,Slaughter2002,Weinberger2005,Dasgupta2013}. 
In particular, we use the expressions provided in Ref.\cite{Dasgupta2013}.

We consider a particular case of unconstrained shear
strain applied to the initially circular inclusion:
\begin{eqnarray}
\epsilon_{\alpha\beta}^*=\epsilon^*\left(2\hat{n}_{\alpha}\hat{n}_{\beta} - \delta_{\alpha\beta}\right)\;,
\label{eq:epsilonP01}
\end{eqnarray}
where $\bm{\hat{n}}$ is a 2-dimensional unit vector that
determines the ``direction" of deformation:
\begin{eqnarray}
\hat{n}_x = \cos(\psi_{IJ})\;,\;\;\;\; \hat{n}_y =\sin(\psi_{IJ})\;.\;\;\;\;\;\;\label{eq:nxnycossin}
\end{eqnarray}

The expression for the final stress field in the inclusion 
(in the absence of external driving force)
from formula (14) of Ref.\cite{Dasgupta2013} is:
\begin{eqnarray}
\sigma_{\alpha\beta}^I=g\epsilon_{\alpha\beta}^*\;,\;\;\;\;\;g\equiv \frac{-\mathcal{E}}{4(1-\nu^2)}\;,\;\;\;
\label{eq:sigmaP01}
\end{eqnarray}
where $\mathcal{E}$ is the Young's modulus, while $\nu$ is the Poisson's ratio.
The eigenvalues and eigenvectors of the stress tensor (\ref{eq:sigmaP01}) can be easily found:
\begin{eqnarray}
&&\lambda_I^1 = +g\epsilon^*,\;\;
V^1_I =\left[\;+\cos(\psi_{IJ}),\;+\sin(\psi_{IJ})\;\right],\;\;\;\label{eq:lamdaVI1}\\
&&\lambda_I^2 = -g\epsilon^*,\;\;
V^2_I =\left[\;-\sin(\psi_{IJ}),\;+\cos(\psi_{IJ})\;\right].\;\;\;\;\;\;\;\label{eq:lamdaVI2}
\end{eqnarray}

The expression for the final stress field in the matrix, according to 
formula (A25) of Ref.\cite{Dasgupta2013}, is:
\begin{eqnarray}
\sigma_{\alpha\beta}^M
=-g\epsilon^*\left\{\left[...\right]_{\alpha\beta}
-4\nu\left(\frac{a^2}{r^2}\right)
\left[\frac{2(\bm{\hat{n}}\bm{r})^2}{r^2}-1\right]\delta_{\alpha\beta}\right\},\label{eq:sigmaPM01}\;\;\;\;\;\;
\end{eqnarray}
where
\begin{eqnarray}
\left[...\right]_{\alpha\beta}&&\label{eq;strainM02}\\
=&&-4(1/\tilde{r})^2
\left\{(1-2\nu)+(1/\tilde{r})^2\right\}\nonumber\\
&&\cdot\left\{\left(\bm{\hat{n}\hat{r}}\right)
\left(\hat{n}_{\alpha}\hat{r}_{\beta}+\hat{n}_{\beta}\hat{r}_{\alpha}\right)
-\hat{r}_{\alpha}\hat{r}_{\beta}\right\}\nonumber\\
&&+(1/\tilde{r})^2\left\{2(1-2\nu)+(1/\tilde{r})^2 \right\}
\left\{2\hat{n}_{\alpha}\hat{n}_{\beta}-
\delta_{\alpha\beta}\right\}\nonumber\\
&&-4(1/\tilde{r})^2\left\{1-2(1/\tilde{r})^2\right\}\left\{2\left(\bm{\hat{n}\hat{r}}\right)^2-
1\right\}\hat{r}_{\alpha}\hat{r}_{\beta}\nonumber\\
&&+4(1/\tilde{r})^2\left\{1-(1/\tilde{r})^2\right\}\nonumber\\
&&\cdot\left\{\left(\bm{\hat{n}\hat{r}}\right)
\left(\hat{n}_{\alpha}\hat{r}_{\beta}+\hat{n}_{\beta}\hat{r}_{\alpha}\right)
-2\left(\bm{\hat{n}\hat{r}}\right)^2\hat{r}_{\alpha}\hat{r}_{\beta}\right\}\nonumber\\
&&+2(1/\tilde{r})^2\left\{1-(1/\tilde{r})^2\right\}\left\{2\left(\bm{\hat{n}\hat{r}}\right)^2 
- 1\right\}\delta_{\alpha\beta}\;.\nonumber
\end{eqnarray}
We are interested in the expression for the stress field in 
the coordinate frame associated with the direction from $I$ to $J$.
In this frame $\bm{\hat{r}}=(1,0)$, while $(\bm{\hat{n}\hat{r}})=\cos(\psi_{IJ})$. 
Also note that $\hat{n}_x \hat{r}_x = \cos(\psi_{IJ})$ and $\hat{n}_y \hat{r}_x = \sin(\psi_{IJ})$, 
while $\hat{n}_x \hat{r}_y = 0$ and $\hat{n}_y \hat{r}_y = 0$.
It is straightforward to obtain from (\ref{eq;strainM02}) the 
following expressions:

\begin{eqnarray}
&&\left[...\right]_{xy}=\left(\frac{a}{r}\right)^2\left\{2-3\left(\frac{a}{r}\right)^2\right\}\sin(2\psi_{IJ})
\;,\;\;\;\;\;\;\label{eq:summxy}\\
&&\left[...\right]_{xx}=\left(\frac{a}{r}\right)^2\left\{-4(1-\nu)+3\left(\frac{a}{r}\right)^2\right\}
\cos(2\psi_{IJ})\;,\;\;\;\;\;\;\;\label{eq:summxx}\\
&&\left[...\right]_{yy}=\left(\frac{a}{r}\right)^2\left\{4\nu-3\left(\frac{a}{r}\right)^2 \right\}
\cos(2\psi_{IJ})\;.\;\;\;\;\;\;\;\;\label{eq:summyy}
\end{eqnarray}

Using expressions (\ref{eq:sigmaPM01}) and (\ref{eq:summxy},\ref{eq:summxx},\ref{eq:summyy}) 
for the stress field in the matrix we get:
\begin{eqnarray}
&&\sigma_{xy}^M(J)=-g\epsilon^*\left(\frac{a}{r}\right)^2
\left\{2-3\left(\frac{a}{r}\right)^2\right\}\sin(2\psi_{IJ})\;,\label{eq:sxyML01}\\
&&\sigma_{xx}^M(J)=-g\epsilon^*\left(\frac{a}{r}\right)^2
\left\{-4+3\left(\frac{a}{r}\right)^2\right\}\cos(2\psi_{IJ})\;,\;\;\;\;\;\;\;\label{eq:sxxML01}\\
&&\sigma_{yy}^M(J)=-g\epsilon^*\left(\frac{a}{r}\right)^2\left\{-3\left(\frac{a}{r}\right)^2\right\}\cos(2\psi_{IJ})\;.\label{eq:syyML01}
\end{eqnarray}
Formulas  (\ref{eq:sxyML01},\ref{eq:sxxML01},\ref{eq:syyML01}) give the components
of the stress tensor at point $J$ in the frame associated with the direction $\bm{r}_{IJ}$.
These stress components are expressed in terms of the magnitude of the inclusion's unconstrained
strain, i.e. $\epsilon^*$, and the direction of the strain, i.e. $\psi_{IJ}$, 
with respect to the direction $\bm{r}_{IJ}$.

The eigenvalues, $\lambda_M^1$ and $\lambda_M^2$, 
and eigenvectors, $\bm{V}^1_M$ and $\bm{V}^2_M$, 
of the stress matrix in the frame associated with
$\bm{r}_{IJ}$ can now be found:
\begin{eqnarray}
&&\lambda^{1}_M=
-g\epsilon^*\left(\frac{a}{r}\right)^2
\left\{-4\left[\cos(\psi_{IJ})\right]^2 + 3\left(\frac{a}{r}\right)^2\right\}\;,\;\;\;\;\;\label{eq:lambda1matrix}\\
&&\lambda^{2}_M=
-g\epsilon^*\left(\frac{a}{r}\right)^2
\left\{+4\left[\sin(\psi_{IJ})\right]^2 - 3\left(\frac{a}{r}\right)^2\right\}\;,\;\;\;\;\;\label{eq:lambda2matrix}\\
&&\bm{V}^1_M = \left[\;+\cos(\psi_{IJ}),\;-\sin(\psi_{IJ})\right]\;,\;\;\;\;\;\;\label{eq:vec1}\\
&&\bm{V}^2_M = \left[\;+\sin(\psi_{IJ}),\;+\cos(\psi_{IJ})\right]\;.\;\;\;\;\;\;\;\label{eq:vec2}
\end{eqnarray}
It follows from (\ref{eq:lamdaVI1},\ref{eq:lamdaVI2}) and (\ref{eq:vec1},\ref{eq:vec2}) that:
\begin{eqnarray}
\psi_{JI}=-\psi_{IJ}\;\;\label{eq:psijiMpsij}.
\end{eqnarray}

Now we are in a position to write expressions for the functions
$\mathcal{F}_1^e$, $\mathcal{F}_2^e$, $\mathcal{F}_3^e$ from
(\ref{eq;calF1e},\ref{eq;calF2e},\ref{eq;calF3e}).
It follows from (\ref{eq:lamdaVI1},\ref{eq:lamdaVI2}) 
that for the inclusion we have:
\begin{eqnarray}
\lambda^1_I - \lambda^2_I = 2g\epsilon^*\;,\;\;\;
\label{eq:lambdaIminus01}
\end{eqnarray}
while for the matrix from (\ref{eq:lambda1matrix},\ref{eq:lambda2matrix}):
\begin{eqnarray}
\lambda^{1}_M - \lambda^{2}_M=-2g\epsilon^*
\left(\frac{a}{r}\right)^2\left\{-2 + 3\left(\frac{a}{r}\right)^2\right\}\;.\;\;\;\;\;\;\;
\label{eq:l1mminusl2m}
\end{eqnarray}
Thus:
\begin{eqnarray}
&&f_o(r)\equiv \left(\lambda^1_I - \lambda^2_I\right)\left(\lambda^{1}_M - \lambda^{2}_M \right)\label{eq:l1mminusl2m}\\
&&=-\left(2g\epsilon^*\right)^2
\left(\frac{a}{r}\right)^2\left\{-2 + 3\left(\frac{a}{r}\right)^2\right\}\;.\;\;\;\;\;
\end{eqnarray}
By taking into account that $\psi_{JI}=-\psi_{IJ}$ from (\ref{eq;calF1e},\ref{eq;calF2e},\ref{eq;calF3e})
we get:
\begin{eqnarray}
\mathcal{F}_1^e = f_o(r) \cos(4\psi_{IJ})\;,\;\;\;
\mathcal{F}_2^e = f_o(r) \;,\;\;
\mathcal{F}_3^e = 0\;.\;\;\;\;\;\;\;\label{eq;calF1eshelby}
\end{eqnarray}
In order to find the average values of the functions above it is necessary to average
them over all values of $\psi_{IJ}$. 
Note that the function $\mathcal{F}_1^e$ averages to zero.
This fact is of interest since in liquids $\langle \mathcal{F}_1^e\rangle $ is not zero and 
it is the function which is the most directly related to viscosity. 
See Fig.\ref{fig:F1F2F5-1} of this paper.



\begin{thebibliography}{10}

\bibitem{EMa20111}
Y.Q. Cheng, E. Ma,
\newblock Progress in materials science {\bf 56}, 379 (2011).

\bibitem{ChenYQ2013}
Y.Q. Cheng, J. Ding and E. Ma,
\newblock Materials Research Letters {\bf 1}, 3 (2013).

\bibitem{Egami19801}
T. Egami, K. Maeda and V. Vitek
\newblock Phil. Mag. A {\bf 41}, 883 (1980). 

\bibitem{Egami19821}
T. Egami and D. Srolovitz,
\newblock J. Phys. F: Met. Phys. {\bf 12}, 2141 (1982). 

\bibitem{Chen19881}
S.P. Chen, T. Egami and V. Vitek, 
\newblock Phys. Rev. B {\bf 37}, 2440 
(1988).

\bibitem{Levashov2008B}
V.A. Levashov, T. Egami, R.S. Aga, J.R. Morris,
\newblock Phys. Rev. B {\bf 78}, 064205 (2008).

\bibitem{Levashov2013}
V.A. Levashov, J.R. Morris, T. Egami,
\newblock J. Chem. Phys. {\bf 138}, 044507 (2013).

\bibitem{Levashov20111}
V.A. Levashov, J.R. Morris, T. Egami,
\newblock Phys. Rev. Lett. {\bf 106}, 115703, (2011).

\bibitem{Egami2007}
T. Egami, S.J. Poon, Z. Zhang, and V. Keppens
\newblock Phys. Rev. B. {\bf 76}, 024203 (2007).

\bibitem{HansenJP20061}
J. P. Hansen and I. R. McDonald, 
\newblock Theory of Simple Liquids, 3rd ed.
Academic Press, London, 2006, Chap. 8.

\bibitem{Boon19911}
J.P. Boon and S. Yip,
\newblock Molecular Hydrodynamics,
Dover Publications Inc., New York, 1991.

\bibitem{Levashov20141}
V.A. Levashov,
\newblock J. Chem. Phys. {\bf 141}, 124502 (2014).

\bibitem{Levashov2014B}
V.A. Levashov,
\newblock Phys. Rev. B {\bf 90}, 174205 (2014).

\bibitem{Bin20151}
B. Wu, T. Iwashita and T. Egami,
\newblock Phys. Rev. E, {\bf 91}, 032301 (2015).

\bibitem{Kust2003a}
T. Kustanovich, Y. Rabin, Z. Olami,
\newblock Phys. Rev. B {\bf 67}, 104206 (2003).

\bibitem{Kust2003b}
T. Kustanovich, Y. Rabin, Z. Olami,
\newblock Physica. A {\bf 330}, 271 (2003).

\bibitem{Levashov2008E}
V.A. Levashov, T. Egami, R.S. Aga, J.R. Morris,
\newblock Phys. Rev. E {\bf 78}, 041202 (2008).  

\bibitem{Plimpton1995}
S. Plimpton, 
\newblock J. Comp. Phys. {\bf 117}, 1-19 (1995).

\bibitem{lammps}
LAMMPS WWW Site:
\newblock http://lammps.sandia.gov.

\bibitem{2Dplots1}
\newblock Figures 3(a), 3(b), and 5(a) of Ref.\cite{Bin20151} can be reproduced using functions
$F_1(r)$, $F_2(r)$, and formula (\ref{eq;ssxy2}) of this paper. Function
$F_3(r)$ that enters into (\ref{eq;ssxy2}) is zero.
\newblock Figures 5(b) and 6(a) of Ref.\cite{Bin20151} can be reproduced using function
$F_5(r)$ and formula (\ref{eq;pxy1}) of this paper. 
Function $F_4(r)$ that enters into (\ref{eq;pxy1}) is zero.
\newblock Figure 5(c) of Ref.\cite{Bin20151} can be reproduced using function
$F_1(r)$ and formula (\ref{eq:sxysdiffcorrf01}) of this paper. 
Function $F_6(r)$ that enters into (\ref{eq:sxysdiffcorrf01}) is zero.
\newblock Figure 6(b) shows function $\langle p_i p_j\rangle $ which is essentially equivalent
to the function $G_{pp}$ from (\ref{eq;Gpp}).
\newblock Figure 6(c) of Ref.\cite{Bin20151} can be reproduced using function
$F_5(r)$ and formula 
$\langle p_i (\sigma^{xx}_{j}-\sigma^{yy}_j)\rangle =F_5(r)\cos(2\theta_{ij})$ 
that can be easily derived.

\bibitem{Eshelby1957}
J. D. Eshelby,
“The Determination of the Elastic Field of an Ellipsoidal Inclusion, and Related Problems”, 
\newblock Proc. R. Soc. Lond. A 241, 376 (1957).

\bibitem{Eshelby1959}
J. D. Eshelby, 
“The Elastic Field Outside an Ellipsoidal Inclusion”, 
\newblock Proc. R. Soc. Lond. A 252, 561 (1959).

\bibitem{Slaughter2002}
W.S. Slaughter,
``The Linearized Theory of Elasticity",
\newblock Springer Science+Business Media New York (2002)

\bibitem{Weinberger2005}
C. Weinberger, W. Cai and D. Burnett, 
``Lecture Notes, Elasticity of microscopic structures",
\newblock http://micro.stanford.edu/~caiwei/me340b\\/content/me340b-notes-v01.pdf

\bibitem{Dasgupta2013}
R. Dasgupta, H.G.E. Hentschel, and I. Procaccia,
\newblock Phys. Rev. E {\bf 87}, 022810 (2013).

\end{thebibliography}
\end{document}